\documentclass[aps,preprintnumbers,twocolumn,amsmath,amssymb,floatfix,pra]{revtex4-1}
\pdfoutput=1
\usepackage[utf8]{inputenc}
\usepackage{amsmath}
\usepackage{amsfonts}
\usepackage{braket}
\usepackage{tensor}
\usepackage{graphicx}
\usepackage[colorlinks=true,linkcolor=blue, citecolor=blue, urlcolor=blue, bookmarks]{hyperref}

\def\be{\begin{equation}}
\def\ee{\end{equation}}
\def\bea{\begin{eqnarray}}
\def\eea{\end{eqnarray}}

\def\XXint#1#2#3{{\setbox0=\hbox{$#1{#2#3}{\int}$}
         \vcenter{\hbox{$#2#3$}}\kern-.5\wd0}}

\begin{document}

\title{Exact dynamics following an interaction quench in a one-dimensional anyonic gas}
\author{Lorenzo Piroli}
\author{Pasquale Calabrese}
\affiliation{SISSA and INFN, via Bonomea 265, 34136 Trieste, Italy.}

\begin{abstract}
We study the nonequilibrium quench dynamics of a one-dimensional anyonic gas. 
We focus  on the integrable anyonic Lieb-Liniger model and consider the quench from non-interacting to hard-core anyons. 
We study the dynamics of the local properties of the system. 
By means of integrability-based methods we compute analytically the one-body density matrix and the density-density correlation function 
at all times after the quench and in particular at infinite time. 
Our results show that the system evolves from an initial state where the local momentum distribution function is non-symmetric to 
a steady state where it becomes symmetric. 
Furthermore, while the initial momentum distribution functions (and the equilibrium ones) explicitly depend on the anyonic parameter, 
the final ones do not. 
This is reminiscent of the dynamical fermionization observed in the context of free expansions after release from a confining trap.
\end{abstract}

\maketitle

\section{Introduction}\label{sec:intro}

The study of the nonequilibrium dynamics of many-body quantum systems has proven to be an effective way of probing important phenomena beyond ground-state and thermal physics. From the experimental point of view, the advent of ultra-cold atoms has provided physicists with unprecedented possibilities of engineering different nonequilibrium settings, shifting the status of several theoretical predictions from purely ideal to actually testable \cite{bdz-08,ccgo-11,pssv-11}. 

One of the most spectacular results of cold-atomic physics has been the possibility of exploring the realm of one-dimensional quantum systems in real experiments. The importance of this breakthrough lies in the prominent role played by dimensionality in many-body quantum physics, as it could be appreciated in connection to the fundamental concept of quantum statistics. Indeed, while in three spatial dimensions particles can be either bosons or fermions, lower dimensional systems allow for anyonic statistics with intermediate properties between the two \cite{lm-77}.
While in two dimensions, anyonic statistics are related to remarkable physical phenomena such as the 
quantum Hall effect \cite{laughlin-83}, one-dimensional anyons have been so far only theoretical speculations. 
Once again, ultra-cold atomic physics is likely to change this situation, with promising experimental schemes already being proposed. In particular, recent works \cite{klmr-11,gs-15, sse-16} suggest that the realization of one-dimensional anyonic gases might be within the reach of current experimental techniques.

Motivated by this general picture, many theoretical studies on one-dimensional anyonic models 
have appeared in the recent literature \cite{agjp-96,rabello-96,it-99,kundu-99,lmp-00,girardeau-06,bgo-06,an-07,pka-07,bgh-07,cm-07,pka-10,sc-09,ssc-07,ghc-09,pka-08,sc-08,hzc-08,pka-09b,patu-15,hao-16,mpc-16}. 
In particular, a detailed analysis of equilibrium properties has been carried out both at zero and finite temperature. Non-trivial theoretical predictions have already been obtained for local correlation functions \cite{bgh-07,cm-07,pka-10,sc-09}, ground-state entanglement \cite{ssc-07,ghc-09}, and the momentum distribution for translationally invariant \cite{pka-08,sc-08,hzc-08,pka-09b,patu-15} and trapped hard-core interacting anyons \cite{Zinn15,hao-16,mpc-16}.

Significantly less attention has been devoted to the study of the nonequilibrium behavior of one-dimensional anyons. Previous theoretical works have mainly focused on protocols where a finite number of particles are released from a confining trap \cite{del Campo-08,hc-12,Li-15} or let evolved after a change in the spatial geometry \cite{wrdk-14}. A conceptually simpler nonequilibrium protocol is that of a quantum 
quench \cite{pssv-11,cc-06,GE15,cem-16}.
For an anyonic gas, a quantum quench is realised by suddenly changing  the interaction between the particles. 
Since this protocol does not require the breaking of translational invariance, it allows us to study in a simple way thermodynamical 
properties of the system out of equilibrium.

In one-dimensional bosonic gases, several interesting and even surprising  aspects of interaction quenches have been investigated over the past decade, as a result of a systematic research activity \cite{cro,grd-10,msf-10,mf-10,ck-12,ia-12,mc-12,ia-12,csc-13,ksc-13,Pozs11-2,m-13,kcc-14,dwbc-14,cgfb-14,
mckc-14,ga-14,cl-14,dc-14,ckc-14b,zwkg-15,th-15,bpc-16,sd-16,dpc-15,bck-15,pce-16,pce2-16,
vwed-16,fgkt-15,pe-16,Bucc16,ccsh-16,bcs-17,dpg-17,dp-16}. In fact, non-trivial features beyond equilibrium physics can been seen already at the level of the post-quench steady state, which is often \emph{qualitatively} different from the equilibrium states of the Hamiltonian driving the time evolution. For example, for quenches to infinitely repulsive interactions the long-time momentum distribution displays a quadratic, rather than quartic, power law decay at large momenta \cite{kcc-14,ksc-13}, thus violating the  Tan's relation, valid quite generally at equilibrium \cite{mvt-02,od-03,tan-08} . Even more interestingly, quenching from repulsive to attractive interactions results, at long times, in thermodynamically stable steady-states displaying either absence of bound states (as in the case of the super-Tonks-Girardeau gas \cite{hgmd-09,abcg-05,bbgo-05,kmt-11,pdc-13}) or an arbitrarily large number of them \cite{pce-16, pce2-16}.

Based on the experience built upon the study of Bose gases, it is certainly intriguing to wonder whether and how anyonic statistics would affect the nonequilibrium dynamics following an interaction quench. As it should be clear from our discussion, the equilibrium case is by now well understood. Indeed, while for one-point functions and several thermodynamic properties a non-vanishing anyonic parameter only results in a renormalization of the pointwise interaction, it leads to dramatic qualitative changes in the behavior of some important observables. This is, for instance, the case of the ground-state  momentum distribution function: for non-zero anyonic parameter it is non-symmetric, signaling the fact that the Hamiltonian breaks parity symmetry \cite{pka-08,sc-08,hzc-08,hao-16,mpc-16}.

In this work we focus on the integrable anyonic Lieb-Liniger model \cite{pka-07,bgh-07} and consider the protocol where the system, initially prepared in the non-interacting ground-state, is quenched to the regime of infinitely large repulsive interactions. In particular, the nonequilibrium dynamics of local correlations after the quench is investigated. Note that the same protocol has been extensively studied in the corresponding bosonic case \cite{grd-10,kcc-14,dc-14,bcs-17,dwbc-14, dpc-15,Bucc16,ccsh-16}. 

Among other results, we compute the time evolution of the anyonic one-body density matrix, which directly yields the momentum distribution function at all times. We discuss the interesting features of the latter, which can be summarized as follows. In the initial state, the momentum distribution is non-symmetric and depends on the anyonic parameter. After a non-trivial post-quench time evolution it approaches a stationary function which, interestingly, is the same for all the anyonic parameters. This is reminiscent of (even though not directly related to) the  dynamical fermionization observed for the free expansion of bosonic \cite{mg-05,rm-05,vxr-17} and anyonic gases \cite{del Campo-08} after release from a confining trap,
as well as in an interaction quench in a bosonic gas \cite{msf-10}.  

The organization of this paper is as follows. In Sec.~\ref{sec:model} we introduce the anyonic Lieb-Liniger Hamiltonian and its Bethe ansatz solution. The quench protocol is explained in Sec.~\ref{sec:quench_protocol} while our results for the full nonequilibrium dynamics are presented in Sec.~\ref{sec:time_ev}. In Sec.~\ref{sec:steady:state} we analyze the post-quench steady state, while Sec.~\ref{sec:computations} contains some technical details of our calculations. Our conclusions are finally reported in Sec.~\ref{sec:conclusions}.
A few appendices contain several technical aspects of our work.

\section{The anyonic Lieb-Liniger model} \label{sec:model}

\subsection{The Hamiltonian and the fields}

We consider the anyonic Lieb-Liniger Hamiltonian which in the formalism of second quantization reads
\begin{multline}
H=\int_{0}^Ldx\left[\partial_x\phi_{\kappa}^{\dagger}(x)\partial_x\phi_{\kappa}(x)\right. \\
+\left.c\phi^{\dagger}_{\kappa}(x)\phi^{\dagger}_{\kappa}(x)\phi_{\kappa}(x)\phi_{\kappa}(x)\right]\,.
\label{eq:hamiltonian}
\end{multline}
Here $\phi_{\kappa}(x)$, $\phi^{\dagger}_{\kappa}(x)$ are anyonic fields with anyonic parameter $\kappa$. They satisfy the generalized commutation relations
\bea
\phi_{\kappa}(x_1)\phi^{\dagger}_{\kappa}(x_2)&=&e^{-i\pi\kappa\epsilon(x_1-x_2)}\phi^{\dagger}_{\kappa}(x_2)\phi_{\kappa}(x_1)\nonumber\\
& &\qquad +\delta(x_1-x_2)\,, \label{eq:CR1}\\
\phi^{\dagger}_{\kappa}(x_1)\phi^{\dagger}_{\kappa}(x_2)&=&e^{i\pi\kappa\epsilon(x_1-x_2)}\phi^{\dagger}_{\kappa}(x_2)\phi^{\dagger}_{\kappa}(x_1)\,,\label{eq:CR2}\\
\phi_{\kappa}(x_1)\phi_{\kappa}(x_2)&=&e^{i\pi\kappa\epsilon(x_1-x_2)}\phi_{\kappa}(x_2)\phi_{\kappa}(x_1)\,,\label{eq:CR3}
\eea
where
\be
\epsilon(x)=\left\{
\begin{array}{cc}
+1\,, & x>0\,,\\
-1\,, & x<0\,,\\
0\,, & x=0\,.
\end{array}\right.
\label{eq:epsilon_function}
\ee
The above relations reduce to traditional bosonic and fermionic commutation relations for $\kappa=0, 1$ respectively. Associated to \eqref{eq:hamiltonian}, one can also define a momentum operator
\bea
P=\frac{i}{2}\int_{0}^Ldx \left[\partial_x\phi_{\kappa}^{\dagger}(x)\phi_{\kappa}(x)-\phi_{\kappa}^{\dagger}(x)\partial_x\phi_{\kappa}(x)\right]\,.
\label{eq:momentum}
\eea

The Hamiltonian \eqref{eq:hamiltonian} generalizes to anyonic particles the well known Lieb-Liniger model \cite{ll-63}. It was introduced and solved using the Bethe ansatz by Kundu \cite{kundu-99}, and systematically analyzed by Batchelor \emph{ et al.} \cite{bgo-06,bgh-07} and P\^atu \emph{et al.} \cite{pka-07,pka-10}. Here we briefly review the aspects relevant to our work, while we refer to the literature for a thorough treatment.

The $N$-particle states  can be generically represented as
\bea
|\chi_N\rangle &=&\frac{1}{\sqrt{N!}}\int_{0}^{L}dx_1\ldots \int_{0}^{L}dx_N \chi_N(x_1,\ldots , x_N)\nonumber\\
&&\times \phi_{\kappa}^{\dagger}(x_1)\ldots\phi_{\kappa}^{\dagger}(x_N)|0\rangle\,,
\label{eq:generic_state}
\eea
where $|0\rangle$ is the Fock vacuum state. In the anyonic case, one has to be careful when imposing boundary conditions. In this work, we will consider periodic boundary conditions on the field operators, namely
\bea
\phi(L)&=&\phi(0)\,,\label{eq:boundary_1}\\
\phi^{\dagger}(L)&=&\phi^{\dagger}(0)\label{eq:boundary_2}\,.
\eea
In contrast to the bosonic and fermionic cases, $\kappa=0, 1$, the wave-function can not be periodic in its coordinates, as a result of consistency between Eqs.~\eqref{eq:boundary_1}, \eqref{eq:boundary_2} and the anyonic commutation relations \eqref{eq:CR1}--\eqref{eq:CR3}. A consistent choice for boundary conditions on the wave-function coordinates  is given by \cite{pka-07}
\bea
\chi_N(0,x_2,\ldots, x_N)&=&\chi_N(L,x_2,\ldots, x_N)\,,\nonumber\\
\chi_N(x_1,0,\ldots, x_N)&=&e^{-i2\pi \kappa}\chi_N(x_1,L,\ldots, x_N)\,,\nonumber\\
&\vdots&\,\nonumber\\
\chi_N(x_1,x_2,\ldots, 0)&=&e^{-i2N\pi \kappa}\chi_N(x_1,\ldots, x_{N-1},L)\,\nonumber.
\eea
Finally, given the commutation relations \eqref{eq:CR1}--\eqref{eq:CR3}, the wave-function in \eqref{eq:generic_state} is well-defined provided that it satisfies the following relation
\bea
\chi_{N}(x_1,\ldots &x_i&,\ldots x_j,\ldots , x_N)=e^{\sigma({\boldsymbol x})}\nonumber\\
&\times &\chi_{N}(x_1,\ldots x_j,\ldots x_i,\ldots , x_N)\,,
\label{eq:CR_wavefunction}
\eea
where
\be
\sigma({\boldsymbol x})\equiv-i\pi\kappa\left[\sum_{k=i+1}^{j}\epsilon(x_i-x_k)-\sum_{k=i+1}^{j-1}\epsilon(x_j-x_k)\right]\,.
\ee

In this paper we will focus on  the one-body density matrix 
\be
\rho_{N}^{\kappa}(x,y,t)=\langle\chi^{0}_{N}|\phi^{\dagger}_{\kappa}(x,t)\phi_{\kappa}(y,t)|\chi^{0}_{N}\rangle\,,
\label{eq:def_rho}
\ee
and on its Fourier transform,  the momentum distribution function 
\be
n_{N}^{\kappa}(q,t)=\frac{1}{L}\int_{0}^{L}dx\int_{0}^{L}dy\rho_{N}^{\kappa}(x,y,t)e^{iq(x-y)}\,.
\label{eq:def_momentum_distribution}
\ee
In particular, we are interested in their thermodynamic limit, which we simply denote with $\rho^{\kappa}(x,y,t)$ and $n^{\kappa}(q,t)$. 

\subsection{The Bethe ansatz solution}\label{sec:bethe_ansatz_solution}

Here, we briefly sketch the Bethe ansatz solution of the Hamiltonian \eqref{eq:hamiltonian} for any value of $c$ and $\kappa$.
As in the well-known bosonic case \cite{ll-63}, each $N$-particle eigenstate of \eqref{eq:hamiltonian} is associated to a set of rapidities $\{\lambda_j\}_{j=1}^N$ which generalize the concept of particle momenta for free Fermi gases. In our case, the rapidities satisfy the following quantization conditions (or Bethe equations)
\be
e^{i\lambda_j L}=e^{i\pi \kappa ( N-1)}\prod_{\substack{k=1\\ k\neq j}}^{N}\frac{\lambda_j-\lambda_k+ic'}{\lambda_j-\lambda_k-ic'}\,,
\label{eq:exp_bethe_eq}
\ee
where
\be
c'=\frac{c}{\cos(\pi\kappa/2)}\,.
\label{eq:coupling}
\ee
In the case of repulsive interactions, the rapidities are real and the Bethe equations \eqref{eq:exp_bethe_eq} can be cast in the convenient logarithmic form
\begin{multline}
\lambda_jL=2\pi I_j-2\sum_{k=1}^{N}\arctan\left(\frac{\lambda_j-\lambda_k}{c'}\right)\\
+2\pi\{\pi\kappa(N-1)\}_{\rm 2\pi}\,,
\label{eq:bethe_equations}
\end{multline}
where the quantum numbers $I_j$ are pairwise distinct, and semi-integer (integer) for $N$ even (odd). Here, following \cite{pka-07}, we also introduced the notation
\be
\{x\}_{2\pi}=\gamma\Leftrightarrow x=2\pi m+2\pi \gamma\,,\quad \gamma\in (-1/2,1/2)\,,
\ee
where $m\in \mathbb{Z}$.

A solution of the Bethe equations \eqref{eq:bethe_equations} is associated to an eigenstate (or Bethe state) with wave-function \cite{pka-07}
\begin{multline}
\chi_N=\frac{e^{-i\pi\kappa/2\sum_{j<k}\epsilon(x_j-x_k)}}{\sqrt{N!\prod_{j>k}\left[(\lambda_j-\lambda_k)^2+c'^2\right]}} \\
\times \sum_{\mathcal{P}\in S_N}(-1)^{\mathcal P}e^{i\sum_{j=1}^Nx_j\lambda_{\mathcal{P}_j}}\mathcal{A}[\{\lambda_j\},\{x_j\}]\,,
\label{eq:eigen_wave_function}
\end{multline}
where
\be
\mathcal{A}[\{\lambda_j\},\{x_j\}]=\prod_{j>k}\left[\lambda_{\mathcal{P}_j}-\lambda_{\mathcal{P}_k}-ic'\epsilon(x_j-x_k)\right]\,,
\ee
and where the sum is over all the permutations $\mathcal{P}$ of the $N$ rapidities. Here $(-1)^{\mathcal{P}}$ denotes the sign of $\mathcal{P}$.
The corresponding energy and momentum are
\bea
E\left[\{\lambda_j\}_j\right]=\sum_{j=1}^{N}\lambda_j^{2}\,,\label{eq:finite_size_energy}\\
P\left[\{\lambda_j\}_j\right]=\sum_{j=1}^{N}\lambda_j\,.
\eea
For an even number of particles, the ground state is obtained by choosing the quantum numbers as
\be
I_j=j-\frac{N+1}{2}\,,\qquad j=1,\ldots N\,.
\ee
Thus, as a consequence of the Bethe equations \eqref{eq:bethe_equations}, the ground state has non-vanishing total momentum
\be
P_0=Np_0=2\pi\{\pi\kappa(N-1)\}_{\rm 2\pi}D\,,
\label{eq:gs_momentum}
\ee
where $D=N/L$ is the density. Note that the momentum per particle $p_0$ vanishes in the thermodynamic limit (defined by $N,L\to\infty $ keeping $D$ fixed), namely
\be
\lim_{\rm th}p_0=0\,.
\label{eq:vanishing_momentum}
\ee

From this discussion, and especially looking at the form of the Bethe equations \eqref{eq:bethe_equations}, it should be clear that the thermodynamics of the anyonic model can be studied along the lines of the corresponding bosonic one as explicitly done in \cite{bgo-06,bgh-07, pka-07,pka-10}. In particular, in the thermodynamic limit the rapidities $\lambda_j$ become continuous variables on the real line following a given distribution function $\rho(\lambda)$. Analogously, one defines the distribution $\rho^{h}(\lambda)$ of holes (i.e. unoccupied states as for the free Fermi gas), and rewrites the Bethe equations \eqref{eq:bethe_equations} in the thermodynamic limit as an integral equation constraining $\rho(\lambda)$ and $\rho^{h}(\lambda)$
\be
\rho(\lambda)+\rho^{h}(\lambda)=\frac{1}{2\pi}+\frac{1}{2\pi}\int_{-\infty}^{\infty}d\mu a(\lambda-\mu)\rho(\mu)\,,
\label{eq:thermo_bethe}
\ee
where
\be
a(\lambda)=\frac{2c'}{\lambda^2+c'^2}\,.
\ee
Note that the form of these equations is the same for all values of the anyonic parameter $\kappa$. The dependence on the latter is implicit through the renormalized coupling~\eqref{eq:coupling}.

\section{The quench protocol}\label{sec:quench_protocol}

In this work we focus on the simplest interaction quench where the coupling is instantaneously changed from $c=0$ to the hard-core limit $c=\infty$. Before turning to the study of the dynamics, we analyze the properties of the initial state and review some features of the hard-core anyonic Hamiltonian governing the time evolution. 

\subsection{The non-interacting ground-state}

For arbitrary anyonic parameter $\kappa$, we choose as initial state  the ground-state of the Hamiltonian \eqref{eq:hamiltonian} 
for vanishing interactions. The corresponding wave-function is obtained from Eq. \eqref{eq:eigen_wave_function} 
taking the limit $c\to 0$. This is non-trivial, as the ground-state rapidities also depend on $c$.
However, in analogy to the small coupling expansion in the bosonic case, we assume that  
for $c\to 0$ the ground-state rapidities satisfy
\be
\lambda_j=p_0+\mathcal{O}\left(\sqrt{c}\right)\,,
\ee
as we verified numerically, for particle numbers up to $N=50$. Here $p_0$ is defined in \eqref{eq:gs_momentum}. 
Using this information, the limit is readily performed, leading to the simple expression
\begin{multline}
\chi^{0}_N(x_1,\ldots , x_n)=\frac{1}{\sqrt{L^N}}e^{-i(\pi \kappa /2)\sum_{j<k}\epsilon\left(x_j-x_k\right)}\\
\times e^{ip_0\left(\sum_{j=1}^Nx_j\right)}\,.
\label{eq:bec_wavefunction}
\end{multline}
The normalized wave-function \eqref{eq:bec_wavefunction} corresponds to our initial state and will be explicitly used in the following for our analytical calculations.

\begin{figure}
\includegraphics[scale=0.78]{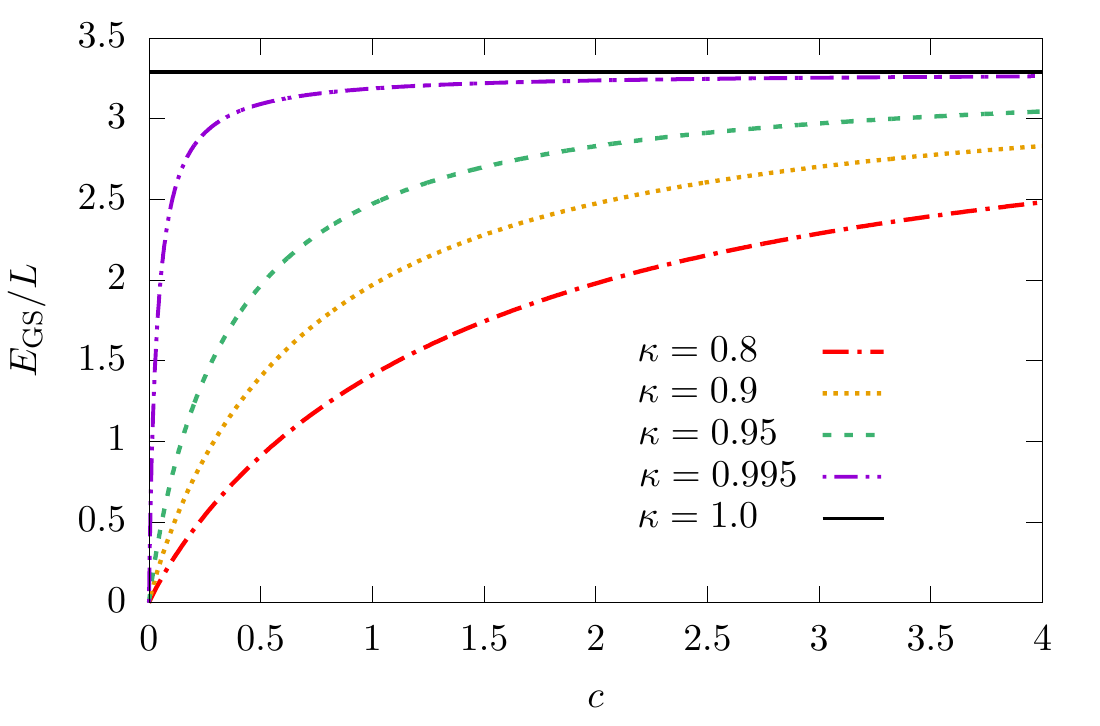}
\caption{Ground-state energy as function of $c$ for $N=L=200$ and different values of $\kappa$. 
The curves are obtained using \eqref{eq:finite_size_energy} and the numerical solution of the Bethe equations \eqref{eq:bethe_equations}. 
It is evident that exchanging the limits $c\to 0$ and $\kappa\to 1$ yields different results for the ground-state energy, thus providing an explicit example of  \eqref{eq:non_commutation_of_limits}.}
\label{fig:1}
\end{figure}

Before proceeding, it is important to note that for the computation of physical quantities the limits $\kappa\to 1$ and $c\to 0$ do not commute. This is evident from the expression of the effective coupling $c'$ in \eqref{eq:coupling}. Indeed
\bea
\lim_{c\to 0} \lim_{\kappa \to 1} c'&=&\infty\,,\\
\lim_{\kappa \to 1} \lim_{c\to 0} c'&=&0\,.
\eea
Hence, for a physical quantity $\mathcal{O}=\mathcal{O}(\kappa,c)$ we will in general have
\bea
\lim_{c\to 0} \lim_{\kappa \to 1} \mathcal{O}(\kappa,c)&\neq & \lim_{\kappa \to 1} \lim_{c\to 0} \mathcal{O}(\kappa,c)\,.
\label{eq:non_commutation_of_limits}
\eea
A simple example is provided by the ground-state energy of the system, as  also reported in \cite{hzc-08}. This is displayed in Fig.~\ref{fig:1}. 

We now turn to the one-body density matrix in the initial state. 
In appendix~\ref{sec:correlations} we briefly discuss some technical aspects regarding the computation of correlation functions on anyonic states. 
From the general expression \eqref{eq:general_correlations}, we have
\bea
\rho_{N}^{\kappa}(x,y,0)&=&N\int_{0}^{L}d^{N-1}z[\chi^{0}_N(x,z_1,\ldots, z_{N-1})]^{\ast}\nonumber\\
&\times &\chi^0_N(y,z_1,\ldots, z_{N-1})\,,
\label{eq:to_evaluate}
\eea
where the wave-function $\chi^{0}_{N}$ is defined in \eqref{eq:bec_wavefunction}.

\begin{figure}
\includegraphics[scale=0.75]{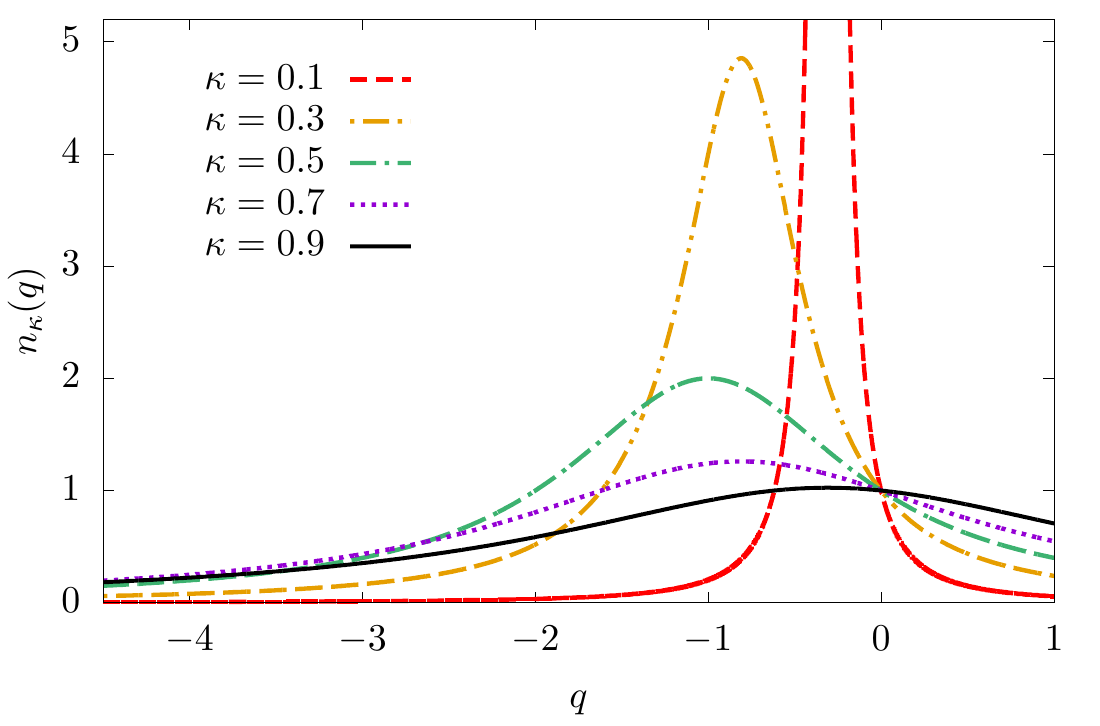}
\caption{Momentum distribution $n_{\kappa}(q)$ in the initial state, for different values of $\kappa$. The plots correspond to the analytical result \eqref{eq:explicit_initial_anyonic_momentum_distribution}. As stressed in the main text, the distributions are clearly not symmetric with respect to $q=0$.}
\label{fig:2}
\end{figure}

Note that the order of the integration variables in \eqref{eq:to_evaluate} is important, as discussed in appendix~\ref{sec:correlations}, and  
that in the calculation, one should always be careful of distinguishing the case $x>y$ or $x<y$. Keeping this in mind, the integrations in \eqref{eq:to_evaluate} can be easily performed, yielding
\begin{multline}
\langle \chi^0_N|\phi_{\kappa}^{\dagger}(x)\phi_{\kappa}(y)|\chi^0_N\rangle= \\
D \Bigg[1-D\frac{|y-x|}{N}\big(1-e^{-\epsilon(y-x)i\pi\kappa}\big)\Bigg]^{N-1}\,,
\label{eq:finite_one_body}
\end{multline}
where $\epsilon(x)$ is the sign function \eqref{eq:epsilon_function}. Taking now the thermodynamic limit of \eqref{eq:finite_one_body}, we arrive at the final result
\bea
\rho^{\kappa}(x,y,0)&=&D\exp\Big\{-D|x-y|\left[1-\nonumber\cos(\pi \kappa)\right]\Big\}\nonumber\\
&\times &\exp\Big\{iD (x-y)\sin(\pi \kappa)\Big\}\,.
\label{eq:initial_OBDM}
\eea
Due to translational invariance, the density matrix $\rho^{\kappa}(x,y,0)$ only depends on the distance $x-y$. This is also true for $t>0$, so that  we can simply define
\be
\rho^{\kappa}(r,t)=\rho^{\kappa}(x+r,x,t)\,.
\ee

For generic $\kappa$ the one-body density matrix \eqref{eq:initial_OBDM} has a non-trivial imaginary part which is the cause of the asymmetry of its Fourier transform. In particular, from \eqref{eq:initial_OBDM} one can immediately compute
\be
n^{\kappa}(q,0)=\frac{2D^2\left[1-\cos(\pi\kappa)\right]}{\left[1-\cos(\pi\kappa)\right]^2D^2+(q+D\sin\kappa\pi)^2}\,,
\label{eq:explicit_initial_anyonic_momentum_distribution}
\ee
which is manifestly asymmetric for $\kappa\neq 0,1$. This can be seen also from Fig.~\ref{fig:2}, where it is plotted for several values of $\kappa$.

Finally, from \eqref{eq:explicit_initial_anyonic_momentum_distribution}, the following limits are straightforwardly computed 
\bea
\lim_{\kappa\to 0}n^{\kappa}(q,0)&=&2\pi D\delta(q)\,,\\
\lim_{\kappa\to 1}n^{\kappa}(q,0)&=&\frac{4D^2}{4D^2+q^2}\,.
\eea
Note that in the limit $\kappa \to 1$ we do not recover a Fermi sea, which would correspond to the ground state of free fermionic particles. This is a manifestation of the fact that the limits $c\to 0$ and $\kappa\to 1$ do not commute. 

\subsection{The hard-core Hamiltonian and the anyon-fermion mapping}
\label{sec:anyon_fermion}

In complete analogy with the case of hard-core bosons, the anyonic gas can be mapped to a free fermionic model in the limit of infinite interactions $c\to \infty$ \cite{girardeau-06}.
First, note that in this limit the wave-function \eqref{eq:eigen_wave_function} takes the simple form (up to a global irrelevant numerical phase)
\bea
\chi_N&=&\frac{e^{-i\pi(\kappa/2)\sum_{j<k}\epsilon(x_j-x_k)}}{\sqrt{N!}}\left[\prod_{j>k}\epsilon(x_j-x_k)\right]\nonumber\\
&\times &\sum_{\mathcal{P}\in S_N}(-1)^{\mathcal P}e^{i\sum_{j=1}^Nx_j\lambda_{\mathcal{P}_j}}\,.
\label{eq:infinitec_wave}
\eea
We recognize that  \eqref{eq:infinitec_wave} is proportional  to a fermionic-wave function: this is the starting point for the explicit mapping between infinitely repulsive anyons and free fermions. This mapping can also be seen at the level of quantum fields, in the language of second quantization. For our convenience, we adopt the latter approach in the following discussion.

For $c\to\infty$, the interacting canonical fields can be thought of as free hard-core anyonic fields, which we denote by $\Phi_{\kappa}$, $\Phi^{\dagger}_{\kappa}$. These satisfy the commutation relations  \eqref{eq:CR1}--\eqref{eq:CR3} but with the additional constraint
\be
\Phi(x)^2=\Phi^{\dagger}(x)^2=0\,.
\ee

The hard-core fields can then be related to fermionic ones $\Psi_F(x)$, $\Psi_F^{\dagger}(x)$, through a generalized Jordan-Wigner mapping \cite{girardeau-06}, which reads
\bea
\Phi_{\kappa}(x)&=&\exp\left[-i\vartheta_{\kappa} \int_0^{x}dz\Psi_F^{\dagger}(z)\Psi_F(z)\right]\Psi_F(x)\,,\label{eq:jordan1}\\
\Psi_F(x)&=&\exp\left[i\vartheta_{\kappa} \int_0^{x}dz\Phi_{\kappa}^{\dagger}(z)\Phi_{\kappa}(z)\right]\Phi_{\kappa}(x)\,,\label{eq:jordan2}
\eea
where
\be
\vartheta_{\kappa}\equiv \pi\left(\kappa-1\right)\,.
\ee
Indeed, using the definition \eqref{eq:jordan2}, it is a simple exercise to show that the fields $\Psi_F(x)$, $\Psi_F^{\dagger}(x)$ satisfy fermionic anticommutation relations
\bea
\{\Psi_F(x),\Psi_F(y)\}&=&0\,,\\
\{\Psi_F(x),\Psi_F^{\dagger}(y)\}&=&\delta(x-y)\,.
\eea
Under this transformation the hard-core anyonic Hamiltonian (equivalent to \eqref{eq:hamiltonian} for $c\to\infty$)
\be
H=\int_{0}^Ldx\ \left[\partial_x\Phi_{\kappa}^{\dagger}(x)\partial_x\Phi_{\kappa}(x)\right]\,,
\label{eq:hardcore_hamiltonian}
\ee
is mapped onto a free fermionic one, so that the operators $\Psi_F(x)$, $\Psi_F^{\dagger}(x)$ evolve in time as free fields.

\begin{figure}
\includegraphics[scale=0.78]{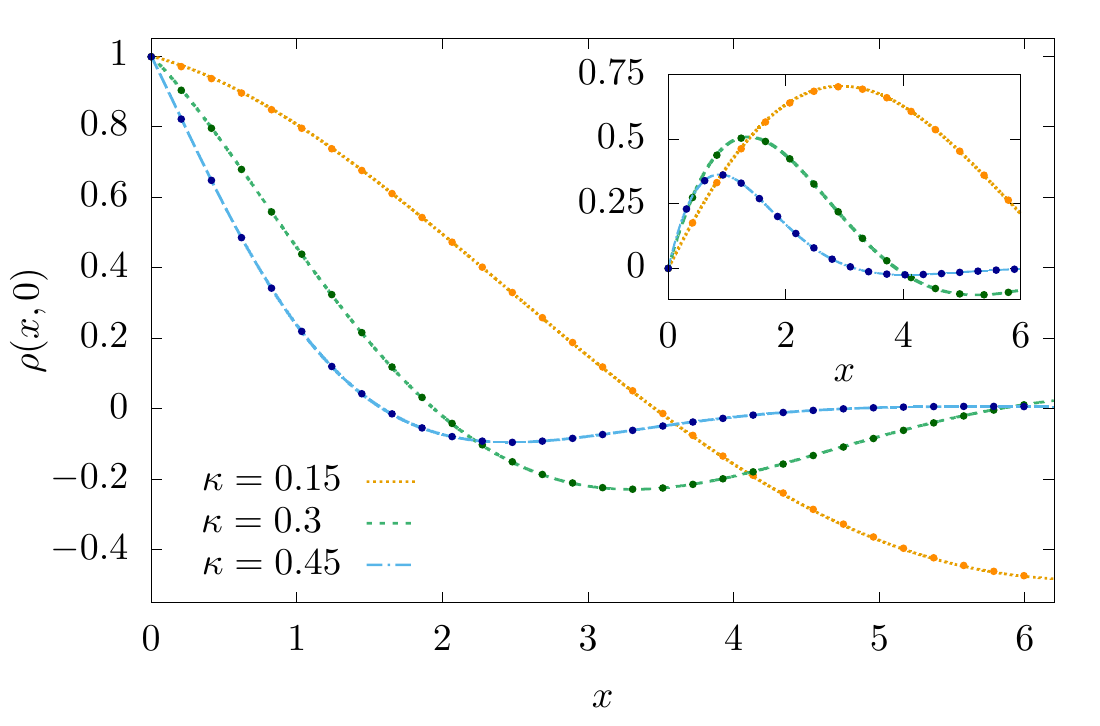}
\caption{Comparison between the analytical result \eqref{eq:initial_OBDM} and the numerical evaluation of \eqref{eq:final_result} for the initial one-body density matrix. Lines correspond to Eq.~\eqref{eq:initial_OBDM} while the dots are the numerical values obtained by evaluating \eqref{eq:final_result} at $t=0$. The main panel and the inset correspond to real and imaginary parts respectively.}
\label{fig:3}
\end{figure}

\begin{figure*}
\includegraphics[scale=0.83]{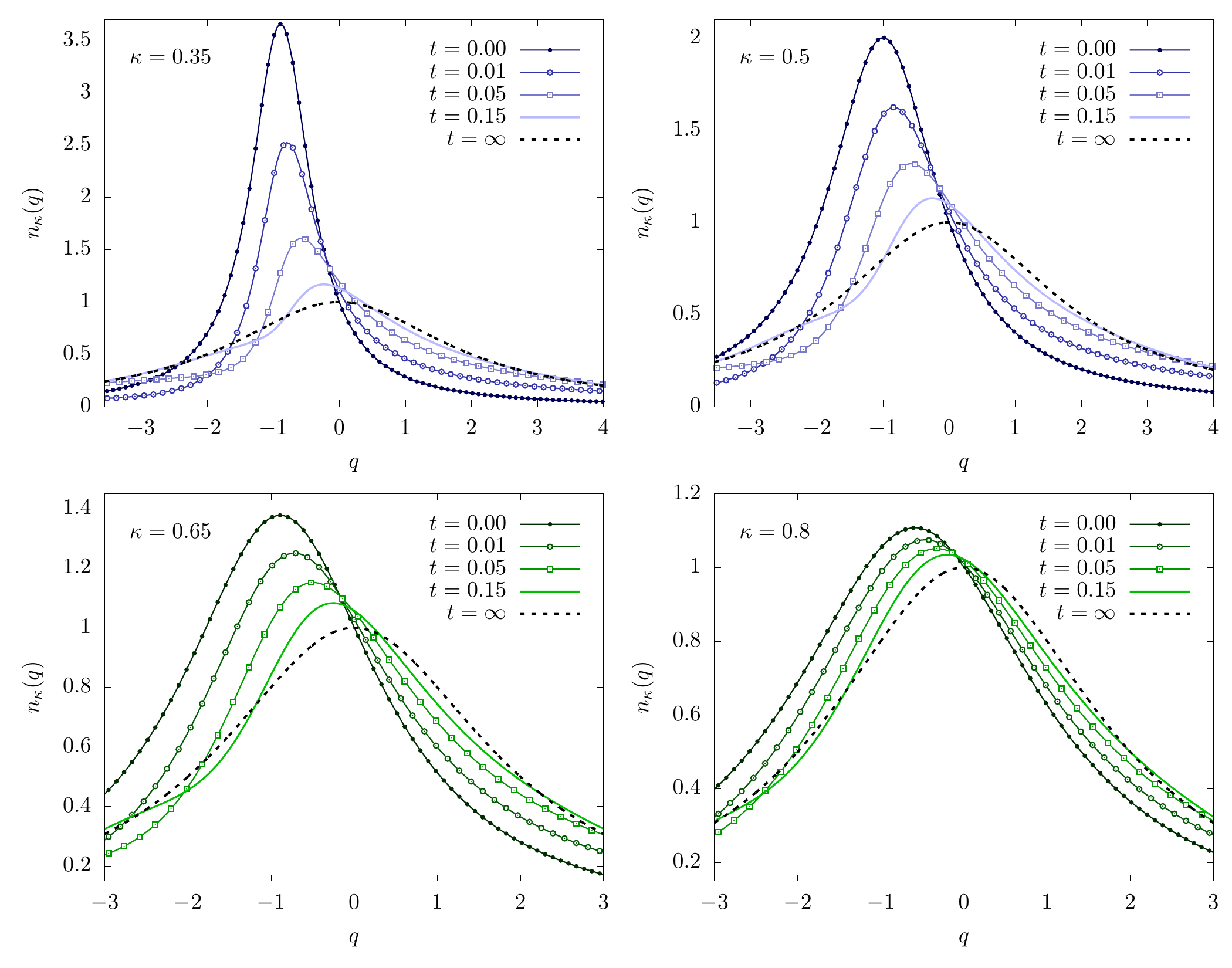}
\caption{Time evolution of the momentum distribution. Continuous colored lines correspond to snapshots at finite times, while the asymptotic steady distribution function \eqref{eq:long_times_momentum} is displayed as a dotted black line. Different plots correspond to different values of the anyonic parameter $\kappa$.}
\label{fig:4}
\end{figure*}

While the equilibrium properties can be then easily worked out in terms of free fermions, 
determining the post-quench  time evolution of anyonic observables remains highly non-trivial, 
due to the non-linear nature of the relations \eqref{eq:jordan1} and \eqref{eq:jordan2}. 
In the bosonic case, analytical results for the time evolution of density-density correlators were first obtained in \cite{kcc-14}, by means of direct calculations involving the Jordan-Wigner transformation. 
The one-body reduced density matrix was computed in \cite{dc-14}, by means of the  Quench Action approach \cite{ce-13,caux-16}, 
because the Jordan-Wigner approach is exceedingly complicated at the technical level.

In this work we will employ both the Quench Action approach and the anyon-fermion mapping: we will use the former to investigate the time evolution of local correlators, while the properties of the post-quench steady state will be best understood by means of the latter.

\section{The post-quench time evolution}\label{sec:time_ev}

In this section we focus on the time evolution of the one-body density matrix and the momentum distribution function: 
these are the most suitable quantities to investigate the anyonic post-quench dynamics, displaying an explicit and non-trivial 
dependence on the anyonic parameter. 
Conversely, density-density correlation functions are independent of $\kappa$, see section~\ref{sec:steady:state}.

Our calculations are based on the Quench Action approach, a method which only relies on some very general structures of Bethe ansatz integrable models \cite{ce-13,caux-16}. In fact, even if previous application were limited to bosonic or fermionic systems, we found that no additional difficulty arises in the presence of anyonic statistics.

Our result builds upon the work \cite{dc-14}, where the quench from non-interacting to hard-core bosons was investigated. By a remarkable calculation, De Nardis and Caux were able to derive \cite{dc-14} a closed-form analytical expression for the time-dependent bosonic density matrix. Despite its conceptual simplicity, the Quench Action approach involves many technical steps. In fact, in order to generalize the derivation of \cite{dc-14} to the anyonic case, several non-trivial calculations are required. For the sake of the presentation, these are postponed to section~\ref{sec:computations},  while here we present and discuss our final result. 

As for the bosonic case \cite{dc-14}, we found that the time-dependent anyonic density matrix is written in terms of two Fredholm determinants, whose definition and properties are briefly reviewed in appendix~\ref{sec:fredholm}. For convenience, in the following will set $D=1$. Then, our final result reads
\begin{widetext}
\bea
\rho^{\kappa}(r,t) &=&  \sqrt{ \text{Det} \begin{pmatrix}
 1 + \mathcal{B}_{\kappa} \rho_{0}  &    \mathcal{B}^{+-}_{\kappa} \varphi_{+}^{(t)} \\   \mathcal{B}_{\kappa}^{+-} \varphi_{-}^{(t)}   & 1  +   \mathcal{B}_{\kappa} \rho_{0}
\end{pmatrix}  } -\sqrt{  \text{Det} \begin{pmatrix}
 1 + \mathcal{A}_{\kappa} \rho_{0}  &  \mathcal{A}^{+-}_{\kappa}  \varphi_{+}^{(t)}\\   \mathcal{A}_{\kappa}^{+-} \varphi_{-}^{(t)}  & 1  +  \mathcal{A}_{\kappa}  \rho_{0}
\end{pmatrix}  } \,.
\label{eq:final_result}
\eea
\end{widetext}
The kernels appearing inside the Fredholm determinants are defined as
\bea
\mathcal{A}_{\kappa}^{+-}(u,v)=\mathcal{A}_{\kappa}(u,-v)\,,\label{eq:A_plusminus}\\
\mathcal{B}_{\kappa}^{+-}(u,v)=\mathcal{B}_{\kappa}(u,-v)\,,
\label{eq:B_plusminus}
\eea
where
\bea
\mathcal{A}_{\kappa}(\lambda,\mu)&=&-\Xi_{\kappa}(r)\frac{2\sin\left[\frac{|r|}{2}(\lambda-\mu)\right]}{\lambda-\mu}\,,\label{eq:a_kernel}\\
\mathcal{B}_{\kappa}(\lambda,\mu)&=&
\mathcal{A}_{\kappa}(\lambda,\mu)+e^{-ir\frac{(\lambda+\mu)}{2}}\,,\label{eq:b_kernel}
\eea
with
\bea
\Xi_{\kappa}(r)&=&\left[1+e^{i\epsilon(r)\pi\kappa}\right]\,,
\label{eq:xi_function}
\eea
and where $\epsilon(x)$ is the sign function \eqref{eq:epsilon_function}.
Finally, in \eqref{eq:final_result} we introduced the functions
\bea
\varphi^{(t)}_+(\lambda)&=&\frac{1}{2\pi}\frac{(\lambda/2)}{1+(\lambda/2)^2}e^{-2it\lambda^2}\,, \label{eq:phi_plus}\\
\varphi^{(t)}_-(\lambda)&=&\frac{1}{2\pi}\frac{(\lambda/2)}{1+(\lambda/2)^2}e^{+2it\lambda^2}\,,\label{eq:phi_minus}\\
\rho_{\rm 0}(\lambda)&=&\frac{1}{2\pi}\frac{1}{1+(\lambda/2)^2}\,.
\label{eq:rho0_function}
\eea
Note that in principle there is an ambiguity for the sign of their square roots in \eqref{eq:final_result}. As we discuss in detail in appendix~\ref{sec:fredholm}, this is not the case and the sign of the square roots is fixed by imposing $\rho^{\kappa}(0,t)=1$, and requiring regularity of $\rho^{\kappa}(r,t)$ as a function of $r$ and $t$.

\begin{figure*}
\includegraphics[scale=0.84]{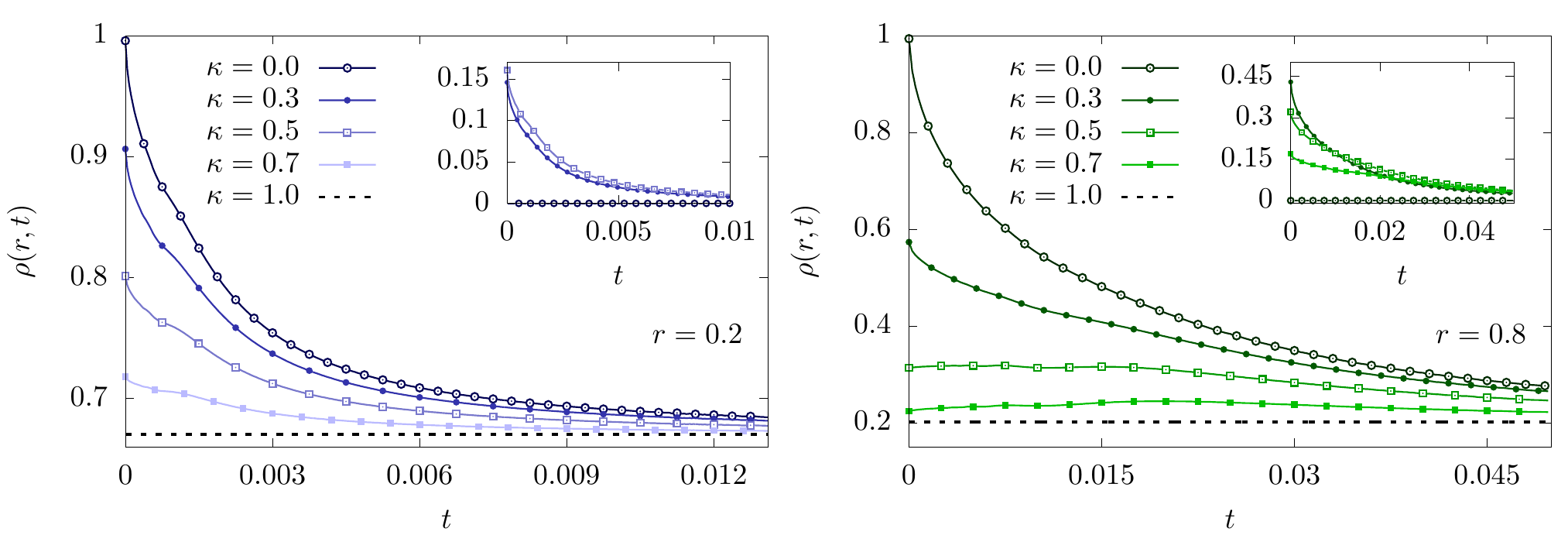}
\caption{Time evolution of the one-body density matrix $\rho^{\kappa}(r,t)$ for different values of the anyonic parameter $\kappa$ and fixed values of $r$. The latter is chosen to be $r=0.2$ (left plot) and $r=0.8$ (right plot). The one-body density matrix is in general complex: in each plot the corresponding real part is reported in the main panel, while the inset displays the imaginary one.}
\label{fig:5}
\end{figure*}

Eq. \eqref{eq:final_result} is particularly interesting because the dependence on the anyonic parameter enters in a very simple way. Indeed, the only modification to the bosonic limit $\kappa=0$ consists in a deformation of the Fredholm kernels, while the structure of the result is the same. Yet, due to non-linearity, several features of the anyonic one-body density matrix are qualitatively different from the bosonic one.

As we discuss in appendix~\ref{sec:fredholm}, the Fredholm determinants in \eqref{eq:final_result} can be easily numerically evaluated using the techniques of \cite{bornemann-10}. A non-trivial check on the validity of \eqref{eq:final_result} is provided by its evaluation at time $t=0$ which has to match the analytical result \eqref{eq:initial_OBDM} (derived by independent methods). We found that the value predicted by \eqref{eq:initial_OBDM} is always reproduced within our numerical precision. This is displayed in Fig.~\ref{fig:3}, showing that the two results are indistinguishable at the scale of the plots. We refer to appendix~\ref{sec:fredholm} for further details on our approach to the evaluation of \eqref{eq:final_result}, and for a discussion on the corresponding numerical precision.

It is natural to consider the limits $\kappa\to 0$, $\kappa\to 1$ of \eqref{eq:final_result}. In the bosonic limit $\kappa\to 0$, it is not difficult to show that \eqref{eq:final_result} is equivalent to the result of \cite{dc-14}, as it should. On the other hand, for $\kappa\to 1$ we explicitly show in appendix~\ref{sec:fermionic_limit} that the r.h.s. of \eqref{eq:final_result} does not depend on time. This is expected: when $\kappa\to 1$ the anyonic fields become fermionic and thus insensitive to repulsive pointwise interactions. Therefore there is no quench to bring the system out of equilibrium.

Inspection of \eqref{eq:final_result} at finite times reveals an extremely interesting behavior. As it is evident from Fig.~\ref{fig:3}, the one-body density matrix is in general complex for $\kappa\neq 0, 1$. The information on its real and imaginary parts is conveniently encoded in its Fourier transform, the momentum distribution, which is always real. In the following we discuss its interesting properties, referring once again to appendix~\ref{sec:fredholm} for its numerical evaluation based on Eq.~\eqref{eq:final_result}.

The time-dependent momentum distribution $n_{\kappa}(q,t)$ is reported in Fig.~\ref{fig:4}. 
The most prominent feature is that the latter evolves from a function which is not symmetric with respect to $q=0$ to a symmetric one. 
Even more interestingly, the asymptotic distribution is the same for every value of the anyonic parameter $\kappa$: at large times the system displays a loss of memory of its anyonic nature.

In order to prove the previous statements, we analytically compute the limit of \eqref{eq:final_result} for infinite times. Since the derivation is rather involved, we reported it in appendix~\ref{sec:long_times}. Conversely, the final result is very simple and reads 
\be
\lim_{t\to\infty}\rho^{\kappa}(x,y,t)=De^{-2|x-y|D}\,,
\label{eq:long_times_rho}
\ee
which immediately yields, after Fourier transform,
\be
\lim_{t\to\infty}n^{\kappa}(q,t)=\frac{4D^2}{q^2+4D^2}\,,
\label{eq:long_times_momentum}
\ee
where we have restored the explicit dependence on the density $D$. This result coincides with the one previously obtained in \cite{ksc-13, kcc-14} for the bosonic case. Our calculations show that this asymptotic behavior is actually generally valid for all the values of the anyonic parameter $\kappa<1$.

The dynamical loss of anyonic memory has been observed before in the literature. Indeed, this was reported by del Campo in \cite{del Campo-08}, where the release of a finite number of hard-core anyons from a confining trap was considered.  Among other results, it was shown in \cite{del Campo-08} that the momentum distributions of confined hard-core repulsive anyons evolve, after release, towards the one of a non-interacting Fermi gas. Accordingly, any dependence on the anyonic parameter is lost at long times.

The work \cite{del Campo-08} builds upon the study of the equivalent protocol for bosonic gases, where the term \emph{dynamical fermionization} was coined \cite{mg-05,rm-05}. In particular, the calculations in \cite{del Campo-08} generalize to the anyonic case those of Minguzzi and Gangardt for one-dimensional Bose gases, as reported in \cite{mg-05}. In the latter work dynamical fermionization was explained based on a scaling property of the exact analytical expression for the time-dependent momentum distribution. 

The interaction quench studied in our work differs in a number of ways from the expansion protocol of \cite{mg-05,rm-05,del Campo-08} (see also \cite{wrdk-14,vxr-17}). Indeed, in our setting translational invariance is not broken and an infinite number of particles is considered. Accordingly, a different mechanism seems to be at the basis of the dynamical loss of anyonic memory in our case. We will see in the next section that this is due to the combination of two concomitant effects occurring in our particular quench protocol. The question of whether this phenomenon might hold for more general interaction quenches remains open, as also discussed in the following.

\begin{figure*}
\includegraphics[scale=0.84]{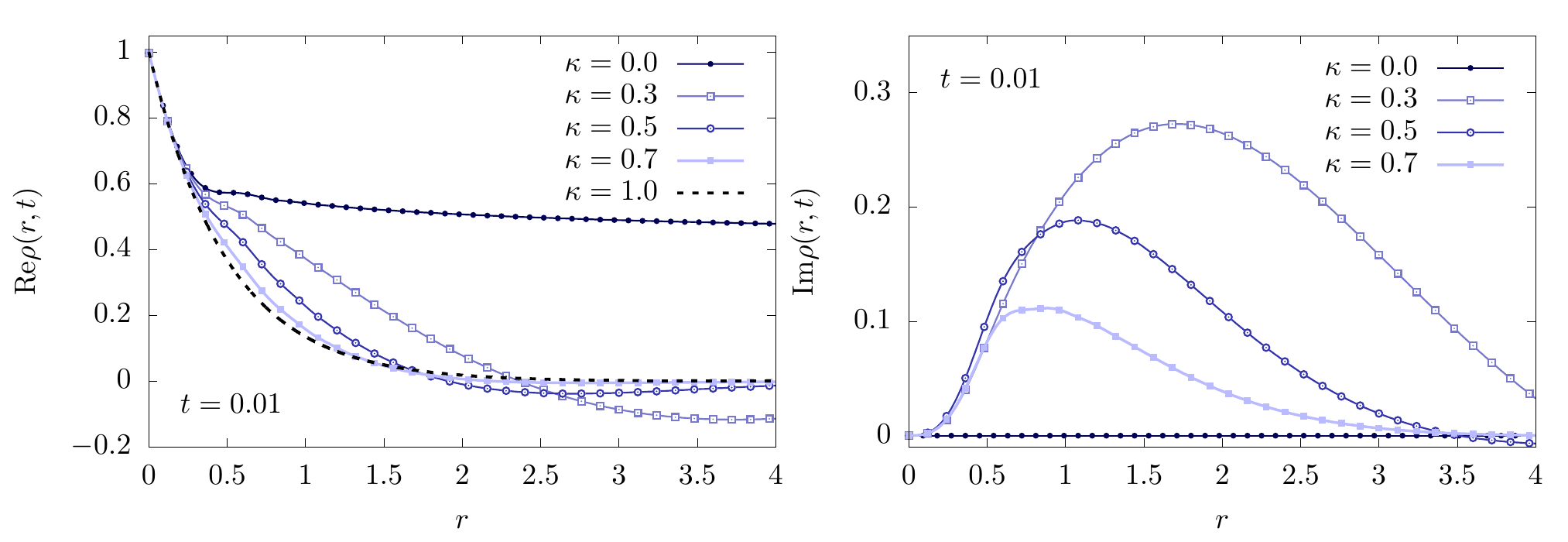}
\caption{One-body reduced density matrix $\rho^{\kappa}(r,t)$ as a function of $r$ for different values of the anyonic parameter $\kappa$ and a fixed value of time. The latter is chosen to be $t=0.01$. The real and imaginary parts of the density matrix are displayed in the left and right plot respectively.} 
\label{fig:6}
\end{figure*}

Before turning to the discussion of the properties of the steady state it is worth to briefly examine the behavior of the two-point function in real-space,
i.e. the one-body reduced density matrix. Its time evolution is reported in Figs. \ref{fig:5} and \ref{fig:6}. 
In Fig. \ref{fig:5} we show the dependence of $\rho^\kappa(r,t)$ on time for two fixed values of the distance, 
while in Fig. \ref{fig:6} we fix the time and show how $\rho^\kappa(r,t)$ varies with the separation $r$.
The first property to mention is that for large time the imaginary part goes always to zero, reflecting the dynamical fermionization 
seen for the momentum distribution. 
However, the most striking feature which should be noticed is the absence of the light-cone spreading of correlations, 
i.e. the fact that connected correlation functions start evolving only after a finite time determined by the maximum velocity of excitations
(see e.g. \cite{cc-06,cetal-12,ef-16,CC:review,kctc-17}). 
This does not come unexpected and indeed it has been also observed for the bosonic counterpart \cite{kcc-14}.
The reason for this unusual phenomenon is that there is not a maximum allowed speed for propagation of signals 
(the mode dependent velocity is proportional to the momentum) and concomitantly the energy pumped into the system by this quench is so large to populate significantly all the single-particle modes.

\section{The steady state}\label{sec:steady:state}

In order to better understand the results obtained in the previous section, we now present a different approach for the computation of the stationary correlators at long times after the quench. This is based on the anyon-fermion mapping introduced in section~\ref{sec:anyon_fermion}.

We recall that the key idea of this mapping is that the hard-core anyonic fields can be explicitly related to fermionic ones through the non-linear relations \eqref{eq:jordan1} and \eqref{eq:jordan2}. Since the fermionic fields evolve freely, the corresponding momentum occupation numbers are conserved in time and their asymptotic value can be conveniently computed in the initial state. 

In the following, we make use of the established result that for free fermionic theories the post-quench steady state can be exactly represented by a generalized Gibbs ensemble (GGE) built out of the conserved momentum occupations numbers \cite{RDYO07,CaEF11,ViRi16,ef-16}. The knowledge of the latter, then, allows us to compute the asymptotic values of all local correlators. Indeed, by Wick's theorem, $n$-point correlators in the GGE can be expressed in terms of two-point functions \cite{ef-16,fe-13b}; in turn, the latter are obtained from the momentum occupation numbers by Fourier transform. In conclusion, we can fully characterize the post-quench steady state by computing
\be
\rho^{F}_N(x,y)=\langle\chi^{0}_N|\Psi_F^{\dagger}(x)\Psi_F(y)|\chi^{0}_N\rangle\,.
\label{eq:fermionic_OBDM}
\ee

We now follow Ref.~\cite{kcc-14}, where analogous calculations were performed for the quench from non-interacting to hard-core Bose gases. Our starting point is given by the following formula \cite{grosse-79}
\bea
\exp\Bigg\{&g&\int_{a}^{b}dy \Phi^{\dagger}_\kappa(y)\Phi_\kappa(y)\Bigg\}\nonumber\\
&=&:\exp\left\{\int_{a}^{b}dy \left(e^{g}-1\right)\Phi^{\dagger}_\kappa(y)\Phi_\kappa(y)\right\}:\,,
\label{eq:grosse_theorem}
\eea
where $:\ldots:$ denotes normal ordering. Equation \eqref{eq:grosse_theorem} is well-known to be valid for canonical bosonic fields  \cite{grosse-79}. However, one can show that \eqref{eq:grosse_theorem} holds also for hard-core anyonic fields with $\kappa\neq 0$. Indeed, by means of the commutation relations \eqref{eq:CR1}--\eqref{eq:CR3}, one can straightforwardly show that the l. h. s. and r. h. s. of \eqref{eq:grosse_theorem} yield the same result when applied on the basis vectors
\be
\Phi^{\dagger}_{\kappa}(x_1)\Phi^{\dagger}_{\kappa}(x_2)\ldots \Phi^{\dagger}_{\kappa}(x_N)|0\rangle\,,\qquad x_1<\ldots <x_N\,.
\ee

Using \eqref{eq:grosse_theorem}, together with \eqref{eq:jordan2}, we obtain
\begin{widetext}
\bea
\langle\chi_N^{0}|\Psi_F^{\dagger}(x)\Psi_F(y)|\chi_N^{0}\rangle &=&\sum_{j=0}^{\infty}\frac{[-\Xi(y-x)]^j}{j!}\nonumber\\
&\times &\left(\prod_{s=1}^{j}\int_{x}^{y}dz_s\right)\langle \chi_N^{0}|\Phi_{\kappa}^{\dagger}(x) \Phi_{\kappa}^{\dagger}(z_1)\ldots \Phi_{\kappa}^{\dagger}(z_j)\Phi_{\kappa}(z_j)\ldots \Phi_{\kappa}(z_1) \Phi_{\kappa}(y)|\chi_N^{0}\rangle\,,
\label{aux_1}
\eea
\end{widetext}
where $\Xi(x)$ is defined in \eqref{eq:xi_function}.
We now specify the calculation for $x<y$ and, following \cite{kcc-14}, we proceed by treating the hard-core fields as canonical fields. Defining for later convenience the $n$-point anyonic correlation function
\bea
\Gamma_{x,y}[\{u_r\}^{j}_{r=1}]&=&\langle \chi_N^0|\Phi_{\kappa}^{\dagger}(x) \Phi_{\kappa}^{\dagger}(u_1)\ldots  \Phi_{\kappa}^{\dagger}(u_j)\nonumber\\
&\times &\Phi_{\kappa}(u_j)\ldots  \Phi_{\kappa}(u_1)\Phi_{\kappa}(y)|\chi_N^0\rangle\,,
\eea
we can exploit the general formula \eqref{eq:general_correlations} to compute 
\bea
\Gamma_{x,y}[\{z_r\}^{j}_{r=1}]&=&\frac{N!}{L^{j+1}(N-j-1)!}\nonumber\\
\times \Bigg[1&-&D\frac{(y-x)}{N}\left(1-e^{-i\pi\kappa}\right)\Bigg]^{N-j-1}e^{ip_0(y-x)}\nonumber\\
&\times &\prod_{k=1}^{j}e^{(-i\pi\kappa/2)\left[-\epsilon(x-z_k)+\epsilon(y-z_k)\right]}\,,
\label{eq:general_correlations_2}
\eea
where $p_0$ is defined in \eqref{eq:gs_momentum}. Hence,  using \eqref{eq:vanishing_momentum}, one obtains for large $N$ and $L$
\bea
\left(\prod_{r=1}^{j}\int_{x}^{y}dz_r\right)\Gamma_{x,y}[\{z_r\}^{j}_{r=1}]=\frac{N!}{L^{j+1}(N-j-1)!}\nonumber\\
\times e^{-D(y-x)|(1-e^{-i\pi\kappa})}\left[e^{-i\kappa\pi}(y-x)\right]^j\,.\hspace{1cm}
\label{aux_2}
\eea
Combining \eqref{aux_1} and \eqref{aux_2} and performing similar steps as those reported in \cite{kcc-14}, we arrive at the fermionic two-point function in the thermodynamic limit, which reads
\be
\lim_{\rm th}\langle\chi^{0}_N|\Psi_{F}^{\dagger}(x)\Psi_{F}(y)|\chi^{0}_N\rangle=De^{-2D(y-x)}\,,\quad y>x\,.
\ee
Analogous steps can be carried out for $x>y$, so that one finally obtains
\be
\lim_{\rm th}\langle\chi_N^{0}|\Psi_{F}^{\dagger}(x)\Psi_{F}(y)|\chi_N^{0}\rangle=De^{-2D|y-x|}\,.
\label{eq:final_OBDM}
\ee
The final formula \eqref{eq:final_OBDM} for the fermionic two point function is independent of the anyonic parameter $\kappa$. In fact, we have precisely recovered the result of \cite{kcc-14} for the bosonic case (which corresponds to $\kappa =0$).

Eq.~\eqref{eq:final_OBDM} allows us to compute the asymptotic value of all the fermionic correlators by means of Wick's theorem, which for free theories is always restored at long times \cite{ef-16}. As a result, we are also able to compute the asymptotics of the anyonic one-body density matrix, exploiting its representation in terms of fermionic fields. Indeed, making once again use of the Jordan-Wigner mapping \eqref{eq:jordan1}, we can write
\begin{widetext}
\bea
\lim_{t\to\infty}\langle \Phi_\kappa^{\dagger}(x)\Phi_\kappa(y)\rangle_t&=&\sum_{j=0}^{\infty}\frac{\left[e^{-i(\kappa-1)\pi}-1\right]^j}{j!}\nonumber\\
&\times &\int_{x}^{y}dz_1\ldots \int_{x}^{y}dz_j \lim_{t\to\infty}\langle \Psi_{F}^{\dagger}(x) \Psi_{F}^{\dagger}(z_1)\ldots \Psi_{F}^{\dagger}(z_j)\Psi_{F}(z_j)\ldots \Psi_{F}(z_1) \Psi_{F}(y)\rangle_t\,\,,
\label{eq:long_times}
\eea
\end{widetext}
where we introduced the notation
\be
\langle \mathcal{O} \rangle_t=\lim_{\rm th}\langle \chi^0_N|\mathcal{O}(t)|\chi^0_N\rangle\,.
\label{eq:notation_time_ev}
\ee
We can now explicitly evaluate the r.h.s. of \eqref{eq:long_times}: by systematic application of Wick's theorem, it is proven in appendix~\ref{eq:wick_theorem} that
\bea
\lim_{t\to\infty}\langle \Psi_{F}^{\dagger}(x) \Psi_{F}^{\dagger}(z_1)\ldots \Psi_{F}^{\dagger}(z_j)\Psi_{F}(z_j)\nonumber\\
\ldots \Psi_{F}(z_1) \Psi_{F}(y)\rangle_t\equiv 0\,,\quad j\geq 1\,.
\label{eq:to_prove2}
\eea
Then, we finally arrive at the extremely simple result
\bea
\lim_{t\to\infty}\langle \Phi_\kappa^{\dagger}(x)\Phi_\kappa(y)\rangle_t &=& \lim_{t\to\infty}\langle \Psi_F^{\dagger}(x)\Psi_F(y)\rangle_t\nonumber\\
&=&\langle \Psi_F^{\dagger}(x)\Psi_F(y)\rangle_{t=0}\,.
\label{eq:final_eq_JW_derivation}
\eea
From Eqs.~\eqref{eq:final_eq_JW_derivation} and \eqref{eq:final_OBDM}, we finally recover the asymptotic expression for the one-body density matrix \eqref{eq:long_times_rho} and for its Fourier transform \eqref{eq:long_times_momentum}.

It is now useful to summarize our results in the light of the above discussion. In the previous section we computed the time-dependent anyonic density matrix and the momentum distribution and showed that they become independent of $\kappa$ when $t\to\infty$. While this emerged as the result of an exact calculation, a transparent explanation for this behavior was missing. Conversely, we have seen in this section that the anyon-fermion mapping allows us to understand the latter in terms of two concomitant effects. 
Specifically:
\begin{itemize}
\item the fermionic occupation numbers of the non-interacting ground-state are independent of $\kappa $;
\item in our quench, fermionic and anyonic two-point functions become equal at long times.
\end{itemize}

If either one of these conditions fails with the other being verified, then the final momentum distribution will explicitly depend on the anyonic parameter $\kappa$. However, for a generic quench, both of these conditions are expected to fail at the same time and it is still possible that the final distribution does not depend on $\kappa$. Hence, it remains an open question whether dynamical loss of anyonic memory might be observed in more general interaction quenches.

We conclude this section by stressing the following point. Since the fermionic representation of the initial state does not depend on the anyonic parameter, any fermionic observable will have the same time evolution for every value of $\kappa$. In particular, since the anyonic and fermionic densities coincide
\be
\langle\Phi_{\kappa}^{\dagger}(x)\Phi_{\kappa}(x)\rangle=\langle\Psi_{F}^{\dagger}(x)\Psi_{F}(x)\rangle\,,
\ee
the time-dependent density-density correlation function will be the same for all the values of $\kappa$. An analytical expression for the latter was obtained for $\kappa=0$ in \cite{kcc-14}; according to our discussion the same result holds more generally for any value of the anyonic parameter. Then, since there is no dependence on $\kappa$, we refer the reader to \cite{kcc-14} for the analytical formula of the density-density correlation function, which will not be reported here.
Another important observable which is independent of the anyonic parameter is the entanglement entropy of a single interval, because the 
Jordan-Wigner transformation, in spite of its non-locality, maps an interval into itself. 
Its steady-state expectation value has been determined in \cite{ckc-14b} and its entire time-evolution can be reconstructed with 
the general technique introduced in \cite{ac-16}.

\section{The time-dependent one-body density matrix}\label{sec:computations}

In this technical section, we finally discuss the derivation of Eq.~\eqref{eq:final_result} by means of the Quench Action approach. In order to do so, we first provide a few general details on the method, referring to \cite{caux-16} for a pedagogical introduction.

For a generic observable, the computation of the post-quench unitary time evolution involves a double summation over the Hilbert space, namely
\be
\langle\chi^0_N|\mathcal{O}(t)|\chi^0_N\rangle=\sum_{m,n}\langle m|\mathcal{O}|n\rangle e^{-i(E_n-E_m)t}\,,
\label{eq:double_sum}
\ee
where we indicated with $|n\rangle$, $|m\rangle$ generic eigenstates with energy eigenvalues $E_n$, $E_m$ respectively. In the thermodynamic limit, this representation is prohibitively complicated both for numerical and analytical calculations. 

One main result of the Quench Action method is that a drastic simplification occurs in integrable systems, for a certain class of \emph{weak} operators \cite{caux-16}. For the latter, which include also those considered in this work, the double summation in \eqref{eq:double_sum} can be replaced by a single one, which is over the excitations of an appropriate \emph{representative} eigenstate. In particular, in the thermodynamic limit, Eq.~\eqref{eq:double_sum} can be cast in the form
\bea
\lim_{\rm th}\langle \chi_N^{0} | \mathcal{O}(t) | \chi_N^{0} \rangle&=&\frac{1}{2}
\sum_{ \mathbf{e} } \Big(   e^{ - \delta s_\mathbf{e} -  i \delta \omega_\mathbf{e} t } \langle \rho_{sp} | \mathcal{O} | \rho_{sp} , \mathbf{e} \rangle\nonumber\\
 &+&   e^{ - \delta s^*_\mathbf{e} +  i \delta \omega_\mathbf{e} t } \langle \rho_{sp},\mathbf{e} | \mathcal{O} | \rho_{sp}  \rangle \Big) \,.
\label{eq:general_time_evolution}
\eea
We now explain the individual elements appearing in this formula.

First, $|\rho_{sp}\rangle$ denotes the representative eigenstate of the system that corresponds to the rapidity distribution function $\rho_{sp}(\lambda)$. The latter depends on the initial state and is found by determining the saddle-point of an appropriate functional, the Quench Action. Its precise form depend again on the initial state and in our case reads 
\bea
S_{QA}[\rho]&=&2S[\rho]-\frac{1}{2}S_{YY}[\rho]
\nonumber\\&+&
h\left(\int_{-\infty}^{+\infty}d\lambda\rho(\lambda)-D\right)\,.
\label{eq:quench_action}
\eea
Here $S_{YY}$ is the Yang-Yang entropy,  
\bea
S_{\rm YY}\left[\rho\right]&=&\int_{-\infty}^{\infty}d\lambda\,\Big\{ \rho(\lambda)\log\left[1+\frac{\rho^{h}(\lambda)}{\rho(\lambda)}\right]\nonumber\\
&+&\rho^{h}(\lambda)\log\left[1+\frac{\rho(\lambda)}{\rho^{h}(\lambda)}\right]\Big\}\,.
\label{eq:yang_yang}
\eea
Note that its expression in the anyonic case is the same as in the bosonic one \cite{bgo-06}; later, we will comment on the overall factor $1/2$ appearing in \eqref{eq:quench_action}. 

The other functional in \eqref{eq:quench_action} is obtained as
\be
S[\rho]=-\lim_{\rm th}\frac{1}{L}{\rm Re}\left[\ln\langle \chi_N^0|\rho\rangle\right]\,.
\label{eq:therm_overlap}
\ee
Here $|\chi^0_N\rangle $ is the initial state while $|\rho\rangle$ represents an eigenstate of the system that is described by the distribution $\rho(\lambda)$ in the thermodynamic limit. Finally, the parameter $h$ in \eqref{eq:quench_action} is a Lagrange multiplier introduced to fix the density of particles. Throughout this section we will set for simplicity $D$=1. Then, the representative (or saddle-point) eigenstate is determined by the condition
\be
\frac{\delta}{\delta \rho}S[\rho]\Big|_{\rho=\rho_{sp}}=0\,.
\label{eq:saddle_point_eq}
\ee

The state $| \rho_{sp} , \mathbf{e} \rangle$ in \eqref{eq:general_time_evolution} is a generic eigenstate which is obtained from $|\rho_{sp}\rangle $ by performing a finite number of particle-hole excitations. As it is well-known from the theory of integrable models, these are defined in strict analogy to the case of free Fermi gases \cite{kbi-93}. Specifically, we remind that each eigenstate is associated, through the Bethe equations \eqref{eq:bethe_equations}, to a set $\{I_j\}$ of quantum numbers. Then, particle (hole) excitations are obtained by adding (removing) a finite number of integers from the set $\{I_j\}$. Note that the summation in \eqref{eq:general_time_evolution} is over all the possible excitations over the representative eigenstate.

Finally, it remains to define $\delta \omega_\mathbf{e}$ and $\delta s_\mathbf{e}$ appearing in \eqref{eq:general_time_evolution}. These correspond to the \emph{differential energy} and the \emph{differential overlap} of the excited states, namely 
\bea
\delta \omega_\mathbf{e} &=& E\left[ |\rho_{sp}, \mathbf{e}\rangle\right]-E\left[| \rho_{sp}\rangle \right]\,,\\
\delta s_\mathbf{e} &=&- \lim_{\rm th}\left({\rm Re}\left[\ln\langle \chi^0_N|\rho_{sp}, \mathbf{e} \rangle\right]\right.\nonumber\\
&+&\left.{\rm Re}\left[\ln\langle \chi_N^0|\rho_{sp}\rangle\right]\right)\,,
\eea
where $E[|\rho\rangle]$ denotes the energy corresponding to the state $|\rho\rangle$.

Having introduced all the building blocks of formula \eqref{eq:general_time_evolution}, we see that in order to actually apply it we need
\begin{itemize}
\item the normalized overlap between the initial state and the Bethe states;
\item the matrix elements (or form factors) of the one-body density matrix between Bethe states.
\end{itemize}
Even in the bosonic case, these quantities are in general extremely hard to compute \cite{dwbc-14,Broc14,Slav89,KoKS97,CaCa06,CaCS07,KoMP10,Pozs11,PaCa14,DePa15,PiCa15,PiCa16,PaDe16}. In addition, at finite values of the interaction $c$ excitations have a non-trivial structure and the expressions for the quantities $\delta \omega_\mathbf{e}$ and $\delta s_\mathbf{e}$, even if known explicitly, are complicated. These difficulties are overcome in the limit $c\to \infty$, where overlaps and form factors can be computed by direct integration of the Bethe wave-functions. For the sake of presentation, this is reported in detail in appendix~\ref{sec:finite_size_calculations}, while in the following we simply present the final results of our calculations.

First, it follows from the discussion in appendix~\ref{sec:finite_size_calculations} that the initial state has a non-vanishing overlap only with those Bethe states for which the rapidities are symmetric with respect to $p_0$ [cf. \eqref{eq:gs_momentum}]. 
In formulas, this condition reads 
\be
\{-(\lambda_j-p_0)\}_{j=1}^{N}=\{(\lambda_j-p_0)\}_{j=1}^{N}\,.
\label{eq:condition2}
\ee
This structure is similar to the one in \cite{Bucc16} in another context.  
In the following we denote with $|\{\lambda_j\}_{j=1}^N\rangle$ a normalized Bethe state with rapidities $\lambda_j$. Furthermore, the set $\{\lambda_j\}_{j=1}^{N}$ will always be understood to be ordered as
\be
\lambda_1>\lambda_2>\ldots >\lambda_N\,.
\label{eq:rapidity_ordering}
\ee
Then, for Bethe states for which \eqref{eq:condition2} is verified, the overlap with our initial state reads
\be
\langle \chi^0_N|\{\lambda_j\}_{j=1}^{N}\rangle=\frac{\sqrt{N!}}{\sqrt{L^{N}}}\bigg(\prod_{j=1}^{N/2}\frac{2}{\lambda_j}\bigg)
\bigg(\prod_{j=1}^{N/2}\left[1-\frac{p_0}{\lambda_j}\right]\bigg)^{-1}\,.
\label{eq:finite_overlap}
\ee

Eq.~\eqref{eq:finite_overlap} allows us to immediate compute the extensive part of the logarithm of the overlap. In particular, we get 
\bea
\lim_{c\to\infty}S[\rho]&=&\frac{D}{2}\left(1-\log D\right)+\int_{0}^{\infty}d\lambda\rho(\lambda)s(\lambda)\,,
\label{eq:thermodynamical_overlap}
\eea
with
\bea
s(\lambda)&=&\ln\left(\lambda/2\right)\,.
\label{eq:single_particle}
\eea
From \eqref{eq:single_particle} and the expression of the Yang-Yang entropy \eqref{eq:yang_yang}, we see that the Quench Action \eqref{eq:quench_action} has the exact same form as in the bosonic case \cite{dwbc-14, dc-14}. In particular, note that the factor $1/2$ in front of the Yang-Yang entropy is due to the constraint \eqref{eq:condition2}, in analogy to \cite{dwbc-14}. As a result, the solution of the saddle-point equation \eqref{eq:saddle_point_eq} coincides with the one in the bosonic quench \cite{dc-14}, namely
\be
\rho_{sp}(\lambda)=\rho_{0}(\lambda)\,,
\ee
where $\rho_{0}(\lambda)$ is defined in \eqref{eq:rho0_function} (where we set $D=1$).

Next, we consider the form factors of the anyonic one-body density matrix between $|\rho_{sp}\rangle$ and the excited states $|\rho_{sp},{\boldsymbol{\rm e}}\rangle$. We recall that the latter is obtained from $|\rho_{sp}\rangle$ by a finite number of particle-hole excitations. Following \cite{dc-14}, we indicate the rapidities of the particle excitations with $\{\mu_j^+\}_{j=1}^{m}$ and those of the hole excitations as $\{\mu_j^-\}_{j=1}^{m}$. Since $p_0$ vanishes in the thermodynamic limit [c.f. \eqref{eq:vanishing_momentum}], we see from \eqref{eq:condition2} that the only particle-hole excitations to be included in the sum \eqref{eq:general_time_evolution} are the parity invariant ones, satisfying
\bea
\{\mu^{\pm}_j\}_{j=1}^{m}=\{\mu^{\pm}_j\}_{j=1}^{n}\cup\{-\mu^{\pm}_j\}_{j=1}^{n}\,,
\label{eq:particle_hole_exc}
\eea
with $m=2n$ and where $\mu_j>0$ are mutually distinct real numbers. Then, the form factor of the excited state associated with \eqref{eq:particle_hole_exc} and $|\rho_{sp}\rangle$ is derived in appendix~\ref{sec:finite_size_calculations} and reads
\begin{widetext}
\bea
\langle \rho_{sp}|\phi^{\dagger}_{\kappa}(x)\phi_{\kappa}(y)|\rho_{sp},\{\mu^-_j,-\mu^-_j\to \mu^+_j,-\mu^+_j\}_{j=1}^n\rangle &=&\left[{\rm Det}\left(1+\mathcal{B}_{\kappa}\rho_{0}\right)\det_{i,j=1}^n
\begin{pmatrix}
\mathcal{W}_{\kappa}(\mu_i^{-},\mu_j^{+})&\mathcal{W}_{\kappa}(\mu_i^{-},-\mu_j^{+})\\
\mathcal{W}_{\kappa}(-\mu_i^{-},\mu_j^{+})&\mathcal{W}_{\kappa}(-\mu_i^{-},-\mu_j^{+})\end{pmatrix}
\right. \nonumber\\
&-& \left.{\rm Det}\left(1+\mathcal{A}_{\kappa}\rho_{0}\right)\det_{i,j=1}^n
\begin{pmatrix}
\mathcal{V}_{\kappa}(\mu_i^{-},\mu_j^{+})&\mathcal{V}_{\kappa}(\mu_i^{-},-\mu_j^{+})\\
\mathcal{V}_{\kappa}(-\mu_i^{-},\mu_j^{+})&\mathcal{V}_{\kappa}(-\mu_i^{-},-\mu_j^{+})
\end{pmatrix}\right]\,.
\label{eq:final_ff}
\eea
\end{widetext}
This formula involves the same Fredholm determinants appearing in \eqref{eq:final_result}, together with the determinant of two $2n\times 2n$ matrices. The latter are expressed in terms of the functions $\mathcal{V}_{\kappa}(\lambda,\mu)$, $\mathcal{W}_{\kappa}(\lambda,\mu)$, which are defined as the solution of the integral equations
\bea
\mathcal{V}_{\kappa}(u,v)&+&\int^{\infty}_{-\infty} ds \mathcal{A}_{\kappa}(u,s)\rho_0(s)\mathcal{V}_{\kappa}(s,v)\nonumber\\
&=&\mathcal{A}_{\kappa}(u,v)\,,\\
\mathcal{W}_{\kappa}(u,v)&+&\int^{\infty}_{-\infty} ds \mathcal{B}_{\kappa}(u,s)\rho_0(s)\mathcal{W}_{\kappa}(s,v)\nonumber\\
&=&\mathcal{B}_{\kappa}(u,v)\,,
\eea
where $\mathcal{A}_{\kappa}$, $\mathcal{B}_{\kappa}$ are given in \eqref{eq:a_kernel}, \eqref{eq:b_kernel}.

Finally, in the limit $c\to\infty$ the expressions for $\delta \omega_\mathbf{e}$ and $\delta s_\mathbf{e}$ are extremely simple, as the Bethe equations \eqref{eq:bethe_equations} become equivalent to the quantization conditions of free Fermi gases. In particular, given an excited state $|\rho_{sp},{\boldsymbol{\rm e}}\rangle$ characterized by particle-hole excitations of the type \eqref{eq:particle_hole_exc}, one has
\bea
\delta \omega_\mathbf{e} &=& \sum_{j=1}^{n}\left[ 2\delta \omega(\mu^{+}_j)- 2\delta\omega(\mu^{-}_j)\right]\,,\\
\delta s_\mathbf{e} &=& \sum_{j=1}^{n} \left[\delta s(\mu^{+}_j)- \delta s(\mu^{-}_j)\right]\,,
\eea
with
\bea
\delta\omega(\mu)&=&\mu^2\,,\label{eq:delta_omega}\\
\delta s(\mu)&=&s(\mu)\,,\label{eq:delta_s}
\eea
and where $s(\mu)$ is given in \eqref{eq:single_particle}.

We have now presented all the ingredients to explicitly derive from the general expression \eqref{eq:general_time_evolution} the final result \eqref{eq:final_result}. From here on the derivation closely follows the one of the bosonic case detailed in \cite{dc-14}. Since it is rather technical, we report it in appendix~\ref{sec:derivation_formula_RDM}, where it is presented for completeness.

\section{Conclusions}\label{sec:conclusions}

We have considered the nonequilibrium dynamics following an interaction quench in the integrable anyonic Lieb-Liniger model and focused on the prototypical case where the system is quenched from non-interacting to hard-core anyons. By means of the anyon-fermion mapping and the Quench Action method we have analytically computed the local correlations of the system at all times after the quench. In particular, we have considered the anyonic one-body density matrix and the momentum distribution function, which turned out to display extremely interesting features. Most prominently, we have shown that the latter evolves from a non-symmetric to a symmetric function, which no longer depends on the anyonic parameter.

A similar loss of anyonic memory has been observed in the context of free expansions of anyonic gases after release from a confining trap \cite{del Campo-08}. In contrast to the situation considered in \cite{del Campo-08}, we have argued that in our quench protocol the loss of anyonic memory is due to a combination of two effects. First, the representation of the non-interacting ground state in terms of fermionic fields is manifestly independent on the anyonic parameter. Second, fermionic and anyonic two-point functions are seen to become equal at long times, due to restoration of Wick's theorem.

It remains an open question to understand whether the same behavior can be observed in more general quenches. Indeed, it would be very interesting to extend our calculations to other initial states, or spatial geometries. Several generalizations are already available in the bosonic case, from initial thermal states \cite{bcs-17}, to confinement on finite segments \cite{mckc-14}. Investigation of the anyonic counterparts of these protocols might help to further explore the nonequilibrium dynamics of one-dimensional anyons.

Another direction to investigate, albeit more difficult, regards the case of attractive interactions. In the context of free expansions after release from a confining trap, this case was also considered in \cite{del Campo-08}, where an effect coined \emph{dynamical bosonization} was observed. On the other hand, several studies on the one-dimensional Bose gas have already shown that quenches to the attractive regime display extremely rich physics \cite{hgmd-09,abcg-05,bbgo-05,kmt-11,pdc-13,pce-16, pce2-16}. We plan to go back to these issues in future works.

\section{Acknowledgments}

We thank Giacomo Marmorini for useful discussions. 
PC acknowledges the financial support by the ERC under Starting Grant 279391 EDEQS.

\appendix

\section{Anyonic correlation functions}\label{sec:correlations}

In this appendix we briefly sketch some properties of  multipoint correlation functions of anyonic fields. Using the generic expression \eqref{eq:generic_state} and the commutation relations \eqref{eq:CR1}--\eqref{eq:CR3}, the computation of scalar products and correlation functions of given states can be reduced to integrals involving the corresponding wave-functions. The easiest example is provided by the norm for which, exploiting the exchange property \eqref{eq:CR_wavefunction}, one has
\be
\langle\chi_N|\chi_N\rangle=\int_{0}^{L}dx_1\ldots \int_{0}^{L}dx_N|\chi_N(x_1,\ldots , x_N)|^2\,.
\ee
The calculation of multipoint correlation functions is slightly more involved, but one can derive the following general expression \cite{pka-08}
\begin{multline}
\langle \chi_N| \phi_{\kappa}^{\dagger}(u_1)\ldots \phi_{\kappa}^{\dagger}(u_n)\phi_{\kappa}(v_n)\ldots \phi_{\kappa}(v_1)|\chi_N\rangle \\
= \frac{N!}{(N-n)!} \int_0^{L}d^{N-n}w\chi_N^{\ast}(u_1,\ldots, u_n,w_1,\ldots , w_{N-n}) \\
\times \chi_N(v_1,\ldots, v_n,w_1,\ldots , w_{N-n})\,.
\label{eq:general_correlations}
\end{multline}

Note that the order of the integration variables in the wave functions appearing in \eqref{eq:general_correlations} is important, due to the exchange relations \eqref{eq:CR_wavefunction}. Eq.~\eqref{eq:general_correlations} is at the basis of several analytical computations performed in this work.

\section{Fredholm determinants and numerical evaluations}\label{sec:fredholm}

In this appendix we briefly discuss Fredholm determinants and their numerical evaluation. In particular, we provide some details on our procedure to evaluate the time-dependent one-body density matrix in \eqref{eq:final_result}. We also explain how to numerically compute the corresponding Fourier transform, displayed in Fig.~\ref{fig:4}.

We consider a kernel $K(x,y)$ and a function $\rho(x)$, defined on a given domain $X\subset \mathbb{R}$.  The corresponding Fredholm determinant is then defined as
\bea
{\rm Det}\left(1+P_{X}K\rho P_{X} \right)&=&\sum_{n=0}^{\infty}\frac{1}{n!}\left(\prod_{j=1}^{n}\int_X dz_j \rho(z_j)\right)\nonumber\\
&&\times \det_{i,j=1}^n K(z_i,z_j)\,.
\label{eq:def_determinant}
\eea
Here $1$ and $P_X$ denote respectively the identity operator and the projector on the domain $X$. When the kernel $K$ and the functions $\rho$ are defined on the whole real line, the projectors are omitted.

The infinite series in \eqref{eq:def_determinant} is in general not suitable for numerical evaluation, which is instead conveniently performed by means of the method explained in \cite{bornemann-10}. For kernels defined on the real line, we introduce a cutoff $\Lambda$ and a finite $m$-point discretization of the interval $X_\Lambda=[-\Lambda,\Lambda]$, which we indicate with $\{ x^{\Lambda}_j \}_{j=1}^M$. A suitable choice for the latter is provided by the Gaussian quadrature, which also gives us a set of weights $\{ w^{\Lambda}_j \}_{j=1}^m$ such that
\be
\int_{-\Lambda}^{\Lambda}f(x)dx= \sum_{j=1}^{N}w^{\Lambda}_{j}f(x^{\Lambda}_j)+\epsilon_N\,,
\ee
where $\epsilon_N$ vanishes as $N\to\infty$. Importantly, Gaussian quadrature can be efficiently implemented numerically \cite{pftv-87}. Then, Fredholm determinants can be evaluated by means of the formula \cite{bornemann-10}
\bea
{\rm Det}( 1+  K \rho )  = \lim_{\Lambda\to\infty} \lim_{m \to \infty} d_{\Lambda,m}\,,
\eea
where
\bea
 d_{\Lambda,m}=\det_{i,j=1}^m \left[ \delta_{ij} +K(x^{\Lambda}_i,x^{\Lambda}_j)\rho(x^{\Lambda}_j)w^{\Lambda}_j \right] \,.
\eea

We have used this method to numerically evaluate the main formula \eqref{eq:final_result} at different values of the distance $r$ and the time $t$.
Due to oscillating terms in \eqref{eq:final_result} depending either on $r$ or on $t$, one needs a rather dense discretization in the interval $X_{\Lambda}=[-\Lambda,\Lambda]$ in order to get accurate results. For this reason, one cannot choose arbitrarily large values of $\Lambda$, since $N$ should also increase accordingly. In summary, the numerical inaccuracy associated with this procedure is due both to the introduction of the cutoff $\Lambda$ and the finite number of points used to evaluate the integrals in $[-\Lambda,\Lambda]$. 

\begin{figure}
\includegraphics[scale=0.78]{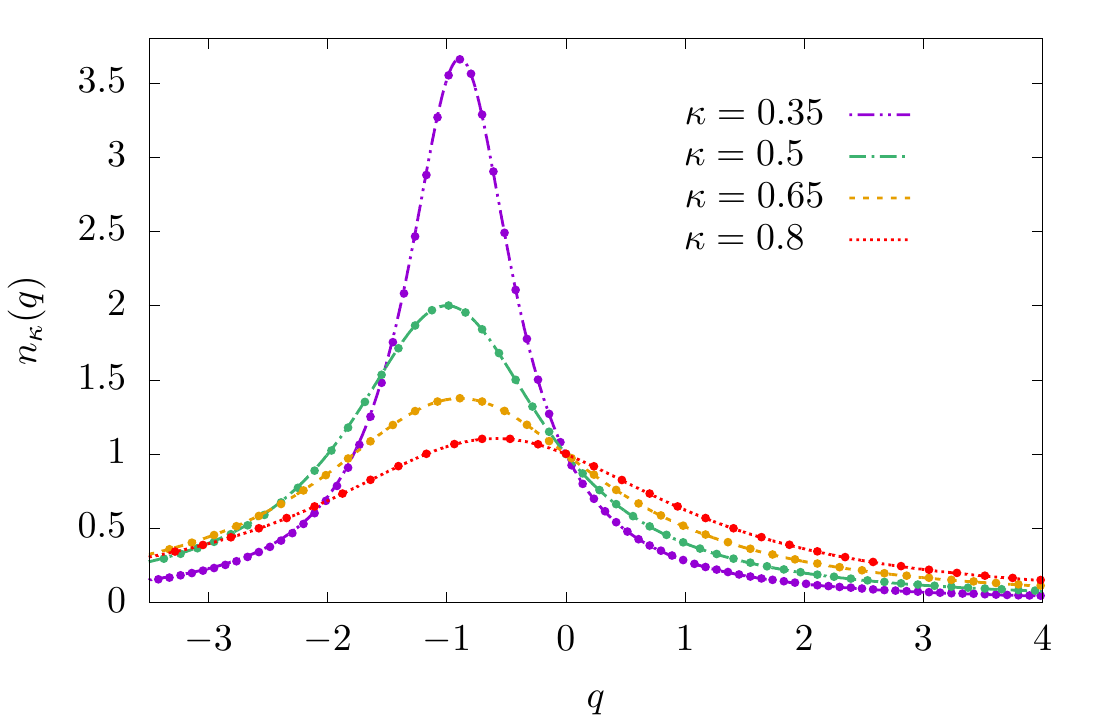}
\caption{Comparison between analytical and numerical results for the initial momentum distribution. Lines correspond to the analytical prediction \eqref{eq:explicit_initial_anyonic_momentum_distribution}. Dots are the numerical values obtained evaluating \eqref{eq:final_result} (at $t=0$) and following the procedure explained in appendix~\ref{sec:fredholm}.}
\label{fig:7}
\end{figure}

In practice, the data displayed in Figs.~\ref{fig:3}--\ref{fig:7} are obtained by choosing values of the cutoff up to $\Lambda=600$, and up to $N=4000$ Gaussian points in $[-\Lambda,\Lambda]$. In order to estimate the precision of our results, we compared the numerical values obtained from \eqref{eq:final_result} at $t=0$ with the analytical prediction \eqref{eq:initial_OBDM}. The absolute error was found to be always smaller than $\epsilon\sim 10^{-3}$. At finite times, the numerical accuracy was estimated by studying variations of the result by increasing $N$ and $\Lambda$, and was again found to be consistent with an absolute error of $\epsilon\sim 10^{-3}$.

It is important to comment about the sign of the square roots of the Fredholm determinants in \eqref{eq:final_result}. 
 In order to unambiguously fix such signs, one requires
\be
\rho^{\kappa}(0,t)=1\,,
\ee
which follows from the fact that the density of particles is conserved after the quench. For $r\neq 0$, then, the signs are fixed accordingly, by requiring regularity of $\rho^{\kappa}(r,t)$ as a function of $r$ and $t$, namely that the latter is continuous with continuous derivatives. These constraints can be easily implemented from the numerical point of view to fix the signs of the square roots. This was explicitly done to produce the data displayed in Figs.~\ref{fig:5} and \ref{fig:6}. Note that choosing the wrong sign in front of either one of the square roots in \eqref{eq:final_result} for some values of $r$ and $t$ would result in abrupt jumps in $\rho^{\kappa}(r,t)$ or in its derivatives, which would be clearly visible in plots like those  in Figs.~\ref{fig:5} and \ref{fig:6}.

We now provide some details on the numerical computation of the time-dependent local momentum distribution function from \eqref{eq:final_result}. For each finite time $t$, we have evaluated the one-body density matrix for a finite number of points $\{r_j\}_{j=1}^M$ in the interval $I_R=[-R,R]$, where $R$ and $M$ have to be chosen sufficiently large. Then, a discrete Fourier transform of the set of values $\{\rho^{\kappa}(r_j,t)\}_{j=1}^M$ was performed. Note that this procedure introduces further sources of numerical error. Indeed, in addition to the inaccuracy associated to the values $\{\rho^{\kappa}(r_j,t)\}_{j=1}^M$ that we already discussed, the Fourier transform is affected by the introduction of the cutoff $R$ and by the approximation of the integrals in $I_{R}$ by discrete sums.

Since the one-body density matrix $\rho^{\kappa}(r,t)$ is exponentially vanishing in $r$, one can take reasonably small values of the cutoff. This was fixed as $R=12$ for the plots in Fig.~\ref{fig:4}, while the we chose $M=256$ discrete points in the interval $I_{R}$. Once again, the absolute numerical error was estimated by comparison of $n^{\kappa}(q,t)$ at $t=0$ with the analytical result \eqref{eq:explicit_initial_anyonic_momentum_distribution}, and was found to be always smaller than $\epsilon\sim 10^{-2}$. The comparison is displayed in Fig.~\ref{fig:7} where we see that the error is invisible on the scale of the plot. Finally, at finite times the numerical accuracy was estimated by studying variations of the result by increasing $M$ and $R$, and was found to be consistent with an absolute error $\epsilon\sim 10^{-2}$.

\section{Long-time limit of the one-body density matrix}\label{sec:long_times}

In this appendix we analytically compute the long-time limit of the one-body density matrix from \eqref{eq:final_result}.

When $t\to\infty$, one can set to zero the rapidly-oscillating off-diagonal terms in the Fredholm determinants appearing in \eqref{eq:final_result}. Then, one immediately obtains 
\be
\lim_{t\to\infty}\rho^{\kappa}(r,t)={\rm Det}\left(1+\mathcal{B}_{\kappa}\rho_{0}\right)- {\rm Det}\left(1+\mathcal{A}_{\kappa}\rho_{0}\right)\,.
\ee
We claim that this expression can be explicitly cast into an extremely simple form, namely
\be
{\rm Det}\left(1+\mathcal{B}_{\kappa}\rho_{0}\right)- {\rm Det}\left(1+\mathcal{A}_{\kappa}\rho_{0}\right)=e^{-2|r|}\,.
\label{eq:to_prove}
\ee
Our claim is well supported by numerical evidence: this equation is always verified within the precision of our evaluation of Fredholm determinants, c.f. appendix~\ref{sec:fredholm}. In addition, we provide in the following further analytical evidence strongly supporting \eqref{eq:to_prove}.

Without loss of generality, we consider the case $r>0$, as an analogous derivation holds when $r<0$.  Note that the kernel $\mathcal{B}_k\rho_{0}$ can be written as the sum of $\mathcal{A}_{\kappa}\rho_{0}$ and a rank--$1$ kernel, namely
\be
\mathcal{B}_{\kappa}\rho_{0}=\mathcal{A}_{\kappa}\rho_{0}+|g\rangle\langle h|\,.
\ee
Here $|g\rangle $ and $|h\rangle$ are complex functions which read
\bea
g(\lambda)&=&e^{-ir\lambda/2}\,, \label{eq:g_vector}\\
h(\lambda)&=&e^{+ir\lambda/2}\rho_{0}(\lambda)\,. \label{eq:h_vector}
\eea

Next, we define
\be
\alpha_{\kappa}=\frac{(1+e^{+i\pi\kappa})}{2}\,,
\ee
so that
\be
\mathcal{A}_{\kappa}(\lambda,\mu)=\alpha_{\kappa}\mathcal{A}_{0}(\lambda,\mu)\,,
\ee
and
\begin{multline}
{\rm Det}\left(1+\mathcal{B}_{\kappa}\rho_{0}\right)- {\rm Det}\left(1+\mathcal{A}_{\kappa}\rho_{0}\right)=\\ 
{\rm Det}( 1 + \alpha_{\kappa} \mathcal{A}_{0} \rho_{0}
+|g\rangle\langle h|)-{\rm Det}\left( 1 + \alpha_{\kappa} \mathcal{A}_{0} \rho_{0}\right)\,.
\end{multline}
Then, introducing the quantity
\bea
\mathcal{D}(\alpha)&=&
{\rm Det}\left( 1 + \alpha \mathcal{A}_{0} \rho_{0}+|g\rangle\langle h|\right)
\nonumber\\&-&{\rm Det}\left( 1 + \alpha \mathcal{A}_{0} \rho_{0}\right)\,,
\eea
we need to prove
\be
\mathcal{D}(\alpha_{\kappa})=e^{-2|r|}\,, \quad \kappa\in [0,1]\,.
\label{eq:to_show3}
\ee
In order to establish \eqref{eq:to_show3}, it is sufficient to show
\bea
\mathcal{D}(0)&=&e^{-2|r|}\,,\label{eq:statement_1}\\
\frac{\partial}{\partial \alpha}\mathcal{D}(\alpha)&=&0\label{eq:statement_2}\,.
\eea

Equation \eqref{eq:statement_1} is immediately derived using the  determinant lemma
\be
{\rm Det}\left(A+|g\rangle\langle h|\right)=\left(1+\langle h|A^{-1}|g\rangle\right){\rm Det A}\,,
\label{eq:determinant_lemma}
\ee
with $A=1$. Indeed, direct calculation gives
\be
1+\langle h|g\rangle=1+e^{-2|r|}\,.
\label{eq:explicit_scalar}
\ee
In order to prove \eqref{eq:statement_2} we use the derivation formula
\be
\frac{\partial }{\partial \alpha}{\rm Det}(A)={\rm Det}(A){\rm tr}\left(A^{-1}\frac{\partial A}{\partial \alpha}\right)\,,
\label{eq:jacobi}
\ee
together with the Sherman-Morrison relation
\be
\left(A+|g\rangle\langle h|\right)^{-1}=A^{-1}-\frac{A^{-1}|g\rangle\langle h|A^{-1}}{1+\langle h|A^{-1}|g\rangle}\,.
\label{eq:sherman_morrison}
\ee
Exploiting \eqref{eq:jacobi} one simply has
\bea
\hspace{-1cm}\frac{\partial }{\partial \alpha}{\rm Det}( &1 &+ \alpha \mathcal{A}_{0} \rho_{0})={\rm Det}\left( 1 + \alpha \mathcal{A}_{0} \rho_{0}\right)\nonumber\\
&\times &{\rm tr}\left[(1 + \alpha \mathcal{A}_{0} \rho_{0})^{-1} \mathcal{A}_{0} \rho_{0}\right]\nonumber\\
&=&\frac{1}{\alpha}{\rm Det}\left( 1 + \alpha \mathcal{A}_{0} \rho_{0}\right){\rm tr}\left[\mathcal{K}_{\alpha}^{-1} (\mathcal{K}_{\alpha}-1)\right]\,,\label{eq:der1}
\eea
where we defined
\be
\mathcal{K}_{\alpha}=1+\alpha \mathcal{A}_0\rho_{0}\,.
\ee
Analogously, exploiting \eqref{eq:jacobi} and \eqref{eq:sherman_morrison},
\begin{widetext}
\bea
\frac{\partial }{\partial \alpha}{\rm Det}( 1 &+& \alpha \mathcal{A}_{0} \rho_{0}+|g\rangle\langle h|)={\rm Det}\left( 1 + \alpha \mathcal{A}_{0} \rho_{0}+|g\rangle\langle h|\right){\rm tr}\left[(1 + \alpha \mathcal{A}_{0} \rho_{0}+|g\rangle\langle h|)^{-1} \mathcal{A}_{0} \rho_{0}\right]\nonumber\\
&=&\frac{1}{\alpha}{\rm Det}\left( 1 + \alpha \mathcal{A}_{0} \rho_{0}\right)(1+\langle h|\mathcal{K}_{\alpha}^{-1}|g\rangle){\rm tr}\left[\left(\mathcal{K}_{\alpha}^{-1}-\frac{\mathcal{K}_{\alpha}^{-1}|g\rangle\langle h|\mathcal{K}_{\alpha}^{-1}}{1+\langle h|\mathcal{K}_{\alpha}^{-1}|g\rangle}\right)(\mathcal{K}_{\alpha}-1)\right]\nonumber\\
&=&\frac{1}{\alpha}{\rm Det}\left( 1 + \alpha \mathcal{A}_{0} \rho_{0}\right){\rm tr}\left\{\left[(1+\langle h|\mathcal{K}_{\alpha}^{-1}|g\rangle)\mathcal{K}_{\alpha}^{-1}-\mathcal{K}_{\alpha}^{-1}|g\rangle\langle h|\mathcal{K}_{\alpha}^{-1}\right](\mathcal{K}_{\alpha}-1)\right\}\,.\label{eq:der2}
\eea
Finally, from \eqref{eq:der1} and \eqref{eq:der2}, using the linearity of the trace, we obtain
\bea
\frac{\partial }{\partial \alpha}{\rm Det}( 1 &+& \alpha \mathcal{A}_{0} \rho_{0}+|g\rangle\langle h|)-
\frac{\partial }{\partial \alpha}{\rm Det}\left( 1 + \alpha \mathcal{A}_{0} \rho_{0}\right)\nonumber\\
&=&\frac{1}{\alpha}{\rm Det}\left( 1 + \alpha \mathcal{A}_{0} \rho_{sp}\right){\rm tr}\left\{\left[\langle h|\mathcal{K}_{\alpha}^{-1}|g\rangle\mathcal{K}_{\alpha}^{-1}-\mathcal{K}_{\alpha}^{-1}|g\rangle\langle h|\mathcal{K}_{\alpha}^{-1}\right](\mathcal{K}_{\alpha}-1)\right\}\,.\label{eq:der3}
\eea
\end{widetext}

Now, the r.h.s. of \eqref{eq:der3} is zero if
\bea
\langle h|\mathcal{K}^{-1}_{\alpha}|g\rangle{\rm tr}\left[\mathcal{K}_{\alpha}^{-1}(\mathcal{K}_{\alpha}-1)\right]\hspace{1cm}\nonumber\\
={\rm tr}\left[\left(\mathcal{K}_{\alpha}^{-1}|g\rangle\langle h|\mathcal{K}_{\alpha}^{-1}\right)(\mathcal{K}_{\alpha}-1)\right]\,.
\label{eq:final_to_prove}
\eea
Introducing the resolvent kernel $R_{\alpha}$ as
\be
1-R_{\alpha}=\mathcal{K}_{\alpha}^{-1}\,,
\label{eq:def_resolvent}
\ee
equation \eqref{eq:final_to_prove} can be rewritten as
\bea
\left(\langle h| g\rangle -\langle h| R_{\alpha}|g\rangle \right){\rm tr} R_{\alpha}&=&\langle h| R_{\alpha}|g\rangle\nonumber\\
&-&\langle h| R_{\alpha}^2|g\rangle\,.
\label{eq:to_prove_resolvent}
\eea
The validity of \eqref{eq:to_prove_resolvent} can finally be established by representing both sides a series expansion in $\alpha$ and comparing each order of the expansion. Indeed, it is easy to derive from the definition \eqref{eq:def_resolvent} the relation 
\be
R_{\alpha}=\alpha \mathcal{A}_{0}\rho_{0}-\alpha \mathcal{A}_{0}\rho_{0}R_{\alpha}\,,
\label{eq:integral}
\ee
from which the following series expansion is immediately derived
\be
R_{\alpha}=\sum_{n=1}^{+\infty}(-1)^{n-1} \alpha^{n}\left(\mathcal{A}_0\rho_{0}\right)^{n}\,.
\label{eq:series_expansion}
\ee
Plugging now \eqref{eq:series_expansion} into \eqref{eq:to_prove_resolvent} one obtains an equality between series expansions in $\alpha$ which can be tested order by order. At each order $n$ one has multiple integrals of simple functions which, at least for small $n$, can be computed analytically. While we were not able to prove the equality \eqref{eq:to_prove_resolvent} for arbitrary order $n$, we analytically checked that the latter holds up to $n=3$. Note that one can in principle extend the analytical checks to higher orders, even though the computations become increasingly unwieldy.

In conclusion, the analytical calculations presented above, together with extensive numerical checks, gives us extremely solid evidence of the validity of \eqref{eq:to_prove}, which we can then safely assume to be true.

\section{The fermionic limit}\label{sec:fermionic_limit}

In this appendix we explicitly show that the result \eqref{eq:final_result} is time-independent in the limit $\kappa\to 1 $. 
%
We start by noticing that the kernels \eqref{eq:a_kernel} and \eqref{eq:b_kernel} greatly simplify for $\kappa=1$, reading
\bea
\mathcal{A}_{\kappa=1} &\equiv& 0\,,\\
\mathcal{B}_{\kappa=1}(\lambda,\mu) &\equiv& e^{-i\frac{r}{2}(\lambda+\mu)}\,.
\eea
Then, using the same notations of appendix~\ref{sec:long_times} we can rewrite
\bea
\mathcal{B}_{\kappa=1}(\lambda,\mu)\rho_0(\mu)=|g\rangle\langle h|\,,\\
\mathcal{B}^{+-}_{\kappa=1}(\lambda,\mu)\varphi^{(t)}_+(\mu)=|g\rangle\langle s^{-}|\,,\\
\mathcal{B}^{+-}_{\kappa=1}(\lambda,\mu)\varphi^{(t)}_-(\mu)=|g\rangle\langle s^{+}|\,.
\eea
The complex functions $|g\rangle $ and $|h\rangle$ were defined in \eqref{eq:g_vector} and \eqref{eq:h_vector}, while $|s^{\pm}\rangle$  correspond to
\bea 
s^{\pm}(\lambda)&=&e^{-ir\lambda/2}\varphi^{(t)}_{\pm}(\lambda)\,.
\eea

Making use of the general formula for the determinant of a block matrix
\bea
 \text{Det} \begin{pmatrix}
A  & B \\ C  & D
\end{pmatrix}  = {\rm Det}(A)  {\rm Det}\left(D-CA^{-1}B\right)\,, \label{eq:det_block}
\eea
we have
\bea
&\ &\text{Det} \begin{pmatrix}
 1 + \mathcal{B}_{\kappa=1} \rho_{0}  &    \mathcal{B}^{+-}_{\kappa=1} \varphi_{+}^{(t)} \\   \mathcal{B}_{\kappa=1}^{+-} \varphi_{-}^{(t)}   & 1  +   \mathcal{B}_{\kappa=1} \rho_{0}
\end{pmatrix}  ={\rm Det}\left(1+|g\rangle\langle h|\right)\nonumber\\
&\times&{\rm Det}\left[1+|g\rangle\langle h|-|g\rangle\langle s^{+}|(1+|g\rangle\langle h| )^{-1}|g\rangle\langle s^{-}|\right]\,.
\eea
From \eqref{eq:sherman_morrison} we obtain
\begin{multline}
|g\rangle\langle s^{+}|(1+|g\rangle\langle h| )^{-1}|g\rangle\langle s^{-}|=\\
|g\rangle\langle s^{+}|  \left(1-\frac{|g\rangle\langle h|}{1+\langle h|g\rangle} \right)|g\rangle\langle s^{-}| = 0\,.
\end{multline}
In the last equality we used the identity
\bea
\langle s^{+}|g \rangle &=&\int_{-\infty}^{\infty}d \lambda e^{i r\lambda/2}\varphi^{(t)}_{+}(\lambda) e^{-i r\lambda/2} =0 \,,
\eea
which follows from the antisymmetry of the functions $\varphi_{\pm}^{(t)}(\lambda)$.

Putting everything together, we finally arrive at
\bea
\rho^{\kappa=1}(r,t) &=&  \text{Det} \left(1+|g\rangle\langle h|\right)-1=e^{-2|r|}\,,
\eea
where the last equality follows from \eqref{eq:determinant_lemma} and \eqref{eq:explicit_scalar}. As previously announced, we see that the one-body density matrix does not depend on time for $\kappa=1$.

\section{Long-time limit of the one-body density matrix from Wick's theorem}
\label{eq:wick_theorem}

In this appendix we prove the validity of Eq.~ \eqref{eq:to_prove2}. Without loss of generality in the following we consider the case $x<y$ (the case $x>y$ is completely analogous), and set $D=1$ for convenience. Furthermore, we introduce the notation
\be
\langle \mathcal{O} \rangle_{\infty}=\lim_{t\to\infty}\lim_{\rm th}\langle \chi_N|\mathcal{O}(t)|\chi_N\rangle\,,
\ee
and denote the fermionic fields simply as $\Psi$, $\Psi^{\dagger}$.
The proof of \eqref{eq:to_prove2} is based on the systematic application of Wick's theorem, which is restored at infinite times as discussed in the main text. First, for $j=1$, applying Wick's theorem and using \eqref{eq:final_OBDM} we obtain
\begin{multline}
 \langle\Psi^{\dagger}(x)\Psi^{\dagger}(z_1)\Psi(z_1)\Psi(y)\rangle_{\infty}=\langle\Psi^{\dagger}(x)\Psi(y)\rangle_{\infty}\\
\times \langle\Psi^{\dagger}(z_1)\Psi(z_1)\rangle_{\infty}
-\langle\Psi^{\dagger}(x)\Psi(z_1)\rangle_{\infty}\langle\Psi^{\dagger}(z_1)\Psi(y)\rangle_{\infty}\\
=0\,,
\end{multline}
where in the last equality we used that by construction $x<z_1< y$. This is true since the integration variable $z_1$ in \eqref{eq:long_times} satisfies $z_1\in (x,y)$.

Next, for $j>1$, we can reorder
\bea
\langle \Psi^{\dagger}(x) \Psi^{\dagger}(z_1)&\ldots& \Psi^{\dagger}(z_j)\Psi(z_j)\ldots \Psi(z_1) \Psi(y)\rangle_{\infty}\nonumber\\
&=& \langle \Psi^{\dagger}(x) \Psi^{\dagger}(w_1)\ldots \Psi^{\dagger}(w_j)\Psi(w_j)\nonumber\\
&\ldots & \Psi(w_1) \Psi(y)\rangle_{\infty}\,,
\label{eq:to_expand}
\eea
where $w_r$ are a reordering of $z_r$ such that $w_1\leq w_2\leq \ldots \leq w_j$.  

Define now the ordered set $\{u_l\}_{l=1}^{j+1}$ by
\be
\left(u_1,u_2,\ldots, u_{j},u_{j+1}\right)=\left(w_1,\ldots, w_j, y\right),
\ee
and expand \eqref{eq:to_expand}, using Wick's theorem, as a sum of $(j+1)!$ terms. Each term is written in the form
\bea
&\ &(-1)^{\mathcal{P}}\langle\Psi^{\dagger}(x)\Psi(u_{\mathcal{P}_1})\rangle_{\infty}\langle\Psi^{\dagger}(w_1)\Psi(u_{\mathcal{P}_2})\rangle_{\infty}  \nonumber\\
&\times& \langle\Psi^{\dagger}(w_3)\Psi(u_{\mathcal{P}_3})\rangle_{\infty}\ldots \langle\Psi^{\dagger}(w_j)\Psi(u_{\mathcal{P}_{j+1}})\rangle_{\infty}\,,
\label{eq:generic_term}
\eea
where $\mathcal{P}$ is an element of the permutation group $S_{j+1}$ of ${j+1}$ elements, and $(-1)^{\mathcal{P}}$ is its sign. Each permutation $\mathcal{P}$ is associated to one and only one permutation $Q$ which is obtained by $\mathcal{P}$ exchanging $u_{\mathcal{P}_1}$ and $u_{\mathcal{P}_2}$. The corresponding term is 
\bea
&-&(-1)^{\mathcal{P}}\langle\Psi^{\dagger}(x)\Psi(u_{\mathcal{P}_2})\rangle_{\infty}\langle\Psi^{\dagger}(w_1)\Psi(u_{\mathcal{P}_1})\rangle_{\infty}\nonumber\\
&\times &\langle\Psi^{\dagger}(w_3)\Psi(u_{\mathcal{P}_3})\rangle_{\infty}\ldots \langle\Psi^{\dagger}(w_j)\Psi(u_{\mathcal{P}_{j+1}})\rangle_{\infty}\,.
\label{eq:generic_term2}
\eea
The sum of \eqref{eq:generic_term} and \eqref{eq:generic_term2} is then vanishing as it is proportional to
\bea
&\ &\langle\Psi^{\dagger}(x)\Psi(u_{\mathcal{P}_1})\rangle_{\infty}\langle\Psi^{\dagger}(w_1)\Psi(u_{\mathcal{P}_2})\rangle_{\infty}\nonumber\\
&\ &\hspace{1cm}-\langle\Psi^{\dagger}(x)\Psi(u_{\mathcal{P}_2})\rangle_{\infty} \langle\Psi^{\dagger}(w_1)\Psi(u_{\mathcal{P}_1})\rangle_{\infty}\nonumber\\
&=& e^{-2|x-u_{\mathcal{P}_1}|-2|w_1-u_{\mathcal{P}_2}|}-e^{-2|x-u_{\mathcal{P}_2}|-2|w_1-u_{\mathcal{P}_1}|}\nonumber\\
&=& e^{-2(u_{\mathcal{P}_1}-x)-2(u_{\mathcal{P}_2}-w_1)}-e^{-2(u_{\mathcal{P}_2}-x)-2(u_{\mathcal{P}_1}-w_1)}\nonumber\\
&=&0\,.
\label{eq:derivation}
\eea
Here we used that by construction $x\leq u_{r}$ and $w_1\leq u_{r}$ for every $r$.

We see that for $j>1$ the quantity in \eqref{eq:to_expand} is written as the sum of pairwise opposite terms, and is thus vanishing. This completes the derivation of \eqref{eq:to_prove2}.

\section{Computation of overlaps and form factors}\label{sec:finite_size_calculations}

In this appendix we compute the normalized overlaps between the initial state and an arbitrary Bethe state for $c\to\infty$, together with the form factors of the anyonic operator $\phi^{\dagger}_{\kappa}(x)\phi_{\kappa}(y)$.

From \eqref{eq:infinitec_wave} we can compute the norm of a Bethe state $|\{\lambda_j\}_{j=1}^{N}\rangle$ as
\bea
\langle\{\lambda_j\}_{j=1}^{N}|\{\lambda_j\}_{j=1}^{N}\rangle &=&\frac{1}{N!}\int_{0}^{L}d^Nx\sum_{P\in S_N}\sum_{Q\in S_N}(-1)^{P+Q}\nonumber\\
&&\times  e^{i\sum_{j=1}^{N}x_j(\lambda_{P_{j}}-\lambda_{Q_{j}})}\,,
\eea
as well as the overlap with the normalized initial state $|\chi^0_N\rangle$ defined by \eqref{eq:bec_wavefunction}
\bea
&\ &\langle \chi^0_N|\{\lambda_j\}_{j=1}^{N}\rangle =\frac{1}{\sqrt{N! L^N}}\int_{0}^{L}d^Nx\sum_{P\in S_N}(-1)^{P}\nonumber\\
&\ &\hspace{1cm}\times e^{i\sum_{j=1}^{N}x_j(\lambda_{P_{j}}-p_0)}\bigg(\prod_{j>k}\epsilon(x_j-x_k)\bigg)\,.
\eea
Introducing the shifted rapidities
\be
\tilde{\lambda}_j=\lambda_j-p_0\,,
\ee
we can rewrite
\bea
\langle\{\lambda_j\}_{j=1}^{N}|\{\lambda_j\}_{j=1}^{N}\rangle &=&\frac{1}{N!}\int_{0}^{L}dx\sum_{P\in S_N}\sum_{Q\in S_N}(-1)^{P+Q}\nonumber\\
&\times & e^{i\sum_{j=1}^{N}x_j(\tilde{\lambda}_{P_{j}}-\tilde{\lambda}_{Q_{j}})}\,,\label{eq:norm}
\eea
and
\bea
&\ &\langle \chi^0|\{\lambda_j\}_{j=1}^{N}\rangle=\frac{1}{\sqrt{N! L^N}}\int_{0}^{L}dx\sum_{P\in S_N}(-1)^{P}\nonumber\\
&\ &\times\bigg(\prod_{j>k}\epsilon(x_j-x_k)\bigg)e^{i\sum_{j=1}^{N}x_j\tilde{\lambda}_{P_{j}}}\,.
\label{eq:overlap}
\eea

Note now that for even $N$, it follows from \eqref{eq:exp_bethe_eq} that in the limit $c\to\infty$ one has
\be
e^{i(\lambda_j-p_0)L}=-1\,,
\ee
namely
\be
e^{i\tilde{\lambda}_jL}=-1\,.
\label{eq:relation}
\ee
Due to the quantization condition \eqref{eq:relation}, the anyonic formulas \eqref{eq:norm}, \eqref{eq:overlap} are now seen to coincide with the corresponding ones in the bosonic case. Then, it is possible to directly exploit the known results for the normalized overlaps between the bosonic non-interacting ground-state and the Bethe states in the limit $c\to \infty$ \cite{dwbc-14}. The only difference consists in the simple substitution $\lambda_j\to\tilde{\lambda}_j$.

As a first result, we then obtain that the overlap is non-vanishing only for those states for which $\{\tilde{\lambda}_j\}=\{-\tilde{\lambda}_j\}$, which is equivalent to \eqref{eq:condition2}. For these states, the overlap formula is \cite{dwbc-14}
\be
\frac{\langle \chi^0_N|\{\lambda_j\}_{j=1}^{N}\rangle}{\langle\{\lambda_j\}_{j=1}^{N}|\{\lambda_j\}_{j=1}^{N}\rangle}=\frac{\sqrt{N!}}{\sqrt{L^{N}}}\left(\prod_{j=1}^{N/2}\frac{2}{\tilde{\lambda}_j}\right)\,,
\label{eq:appendix_finite_size_overlap}
\ee
where the ordering of the rapidities is the same as in \eqref{eq:rapidity_ordering}. Eq.~\eqref{eq:appendix_finite_size_overlap} can be easily cast in the form \eqref{eq:finite_overlap}, where the overlap is rewritten in terms of the rapidities $\lambda_j$.  Note that for parity invariant states, one always has
\be
|\lambda_j-p_0|\geq \frac{2\pi}{L}\,,
\ee
so that the overlap \eqref{eq:finite_overlap} is never singular. Using this information and \eqref{eq:vanishing_momentum}, one can show that the last factor of \eqref{eq:finite_overlap} can be dropped in the thermodynamic limit, so that one simply obtains \eqref{eq:thermodynamical_overlap}.

We now compute the (normalized) form factors of the one-body density matrix between two arbitrary Bethe states for $c\to\infty$. The calculations are along the lines of those presented in \cite{dc-14} for the quench from non-interacting to hard-core bosons. 

First, it is easy to show from \eqref{eq:infinitec_wave} that
\be
\langle\{\lambda_j\}|\{\lambda_j\}\rangle=L^{N}\,.
\label{eq:normalization}
\ee
With a slight abuse of notation, from here on we will denote with $|\{\lambda_j\}\rangle$ normalized states (since the normalization \eqref{eq:normalization} is trivial, this will not generate confusion). In the following, we consider the rapidities $\{\lambda_j\}$ to be arbitrary real numbers, for which one can still formally define a Bethe state through \eqref{eq:infinitec_wave}. Such states are usually called \emph{ off-shell}. Only at the end of the calculation we will take the rapidities $\{\lambda_j\}$ to be a solution of the Bethe equations \eqref{eq:bethe_equations}.

For $y> x$ one can make use of the general formula \eqref{eq:general_correlations} to compute
\bea
&\ &\langle\{\lambda_j\}|\phi_{\kappa}^{\dagger}(x)\phi_{\kappa}(y)|\{\mu_j\}\rangle =\frac{1}{L}\frac{1}{(N-1)!}\sum_{P,Q}(-1)^{P+Q}\nonumber\\
&\times &e^{-ix\lambda_{P_1}+iy\mu_{Q_1}}\prod_{j=2}^N f_{+}(\lambda_{P_j}-\mu_{Q_j})\,,
\label{eq:intermediate}
\eea
where
\bea
f_{+}(\lambda)&=&-\frac{e^{-i\lambda y}}{i\lambda L}\left[e^{-i\lambda (x-y)}-1-e^{-i\pi\kappa}\left(1-e^{-i\lambda(x-y)}\right)\right.\nonumber\\
&+&\left.\left(e^{-i\lambda L}-1\right)e^{i\lambda y}\right]\,.
\eea
After standard manipulations, \eqref{eq:intermediate} can be rewritten as
\bea
&\ &\langle\{\lambda_j\}|\phi_{\kappa}^{\dagger}(x)\phi_{\kappa}(y)|\{\mu_j\}\rangle 
=e_{\lambda,\mu}\frac{1}{L}\sum_{P'}(-1)^{P'}\nonumber\\
&\times &\prod_{j=1}^N q_{+}(\lambda_{j}-\mu_{P'_j})\sum_{k=1}^N\frac{e^{-i(x-y)\lambda_k}}{q_+(\lambda_k-\mu_{P'_k})}\,,
\label{eq:intermediate3}
\eea
where
\bea
e_{\lambda,\mu}&=&\exp\left\{-iy\left[\sum_{j=1}^N\left(\lambda_j-\mu_j\right)\right]\right\}\,,\label{eq:e_factor}
\eea
and
\bea
&\ &q_+(\lambda)=-\frac{1}{i\lambda L}\left[e^{-i\lambda (x-y)}-1-e^{-i\pi\kappa}\right.\nonumber\\
&\times &\left. \left(1-e^{-i\lambda(x-y)}\right)
+\left(e^{-i\lambda L}-1\right)e^{i\lambda y}\right]\,.
\eea
Finally, as in \cite{dc-14} we note that \eqref{eq:intermediate3} can be written as the difference of two determinants, namely
\bea
&\ &\langle\{\lambda_j\}|\phi_{\kappa}^{\dagger}(x)\phi_{\kappa}(y)|\{\mu_j\}\rangle =\nonumber\\
&&e_{\lambda,\mu} \left(\det_N\left[q_{+}(\lambda_a-\mu_b)+\frac{e^{-i\lambda_a(x-y)}}{L}\right]\right.\nonumber\\
&-&\left.\det_N\left[q_{+}(\lambda_a-\mu_b)\right]\right)\,, \quad x<y\,.
\label{eq:final_finite_size_ff}
\eea
An analogous computation can be carried out for $x>y$.

Eq.~\eqref{eq:final_finite_size_ff} allows us to compute the thermodynamic limit of the form factors between the representative eigenstate $|\rho_{sp}\rangle$ and an eigenstate obtained from it by performing a finite number of particle-hole excitations. As in the main text, we indicate the rapidities of the particle excitations as $\{\mu_j^+\}_{j=1}^{m}$ and those of the hole excitations as $\{\mu_j^-\}_{j=1}^{m}$. Note that we are always interested in form factors between Bethe states with the same momentum, so that one can set $e_{\lambda,\mu}=1$ in \eqref{eq:final_finite_size_ff}. Then, it straightforward to repeat the steps outlined in \cite{dc-14} to compute the thermodynamic limit of the form factor \eqref{eq:final_finite_size_ff}, yielding the final result
\bea
&\ &\langle \rho_{sp}|\phi_{\kappa}^{\dagger}(x)\phi_{\kappa}(y)|\rho_{sp},\{\mu^-_j\to \mu^+_j\}_{j=1}^m\rangle =\nonumber\\
&=&\left[{\rm Det}\left(1+\mathcal{B}_{\kappa}\rho_{0}\right)\det_{i,j=1}^m\mathcal{W}_{\kappa}(\mu_i^{-},\mu_j^{+})\right. \nonumber\\
&-& \left.{\rm Det}\left(1+\mathcal{A}_{\kappa}\rho_{0}\right)\det_{i,j=1}^m\mathcal{V}_{\kappa}(\mu_i^{-},\mu_j^{+})\right]\,.
\eea
This formula involves the same Fredholm determinants appearing in \eqref{eq:final_result}, together with the determinant of two $m\times m$ matrices. The latter are expressed in terms of the functions $\mathcal{V}_{\kappa}(\lambda,\mu)$, $\mathcal{W}_{\kappa}(\lambda,\mu)$, which are defined as the solution of the integral equations
\bea
\mathcal{V}_{\kappa}(u,v)&+&\int^{\infty}_{-\infty} ds \mathcal{A}_{\kappa}(u,s)\rho_0(s)\mathcal{V}_{\kappa}(s,v)\nonumber\\
&=&\mathcal{A}_{\kappa}(u,v)\,,\\
\mathcal{W}_{\kappa}(u,v)&+&\int^{\infty}_{-\infty} ds \mathcal{B}_{\kappa}(u,s)\rho_0(s)\mathcal{W}_{\kappa}(s,v)\nonumber\\
&=&\mathcal{B}_{\kappa}(u,v)\,,
\eea
where $\mathcal{A}_{\kappa}$, $\mathcal{B}_{\kappa}$ are given in \eqref{eq:a_kernel}, \eqref{eq:b_kernel}.

We conclude this appendix with an important remark. In the case of non-vanishing anyonic parameter $\kappa$, the form factors are in general complex-valued. However, by repeating the calculations presented in this appendix for $y<x$ one can show that the following property holds
\be
\langle\{\lambda_j\}|\phi_{\kappa}^{\dagger}(x)\phi_{\kappa}(y)|\{\mu_j\}\rangle= \left(\langle\{\lambda_j\}|\phi_{\kappa}^{\dagger}(y)\phi_{\kappa}(x)|\{\mu_j\}\rangle\right)^\ast\,.
\label{eq:ff_exchange_identity0}
\ee
Note that \eqref{eq:ff_exchange_identity0} holds for \emph{on-shell} Bethe states, namely in the case where the rapidities $\{\lambda_j\}$ satisfy the Bethe equations \eqref{eq:bethe_equations}. From \eqref{eq:ff_exchange_identity0}, we can immediately deduce
\bea
\langle \{\mu_j\}|\phi_{\kappa}^{\dagger}(x)\phi_{\kappa}(y)|\{\lambda_j\}\rangle\hspace{0.5cm}\nonumber\\
=  \langle\{\lambda_j\}|\phi_{\kappa}^{\dagger}(x)\phi_{\kappa}(y)|\{\mu_j\}\rangle \,.
\label{eq:ff_exchange_identity}
\eea
This identity will be explicitly used in appendix~\ref{sec:derivation_formula_RDM} for the time evolution of the one-body reduced density matrix.

\onecolumngrid

\section{Computation of the time-dependent one-body density matrix }\label{sec:derivation_formula_RDM}

In this appendix we present for completeness the derivation of the main formula \eqref{eq:final_result} by means of the Quench Action method. As stressed in the main text, the derivation closely follows the one presented in Ref.~\cite{dc-14}, where the quench from non-interacting to hard-core bosons was studied. Given the strict analogies, throughout our derivation we will sometimes refer to Ref.~\cite{dc-14} for further details on technical aspects of the calculations.

The starting point of our derivation is provided by the general formula \eqref{eq:general_time_evolution}, whose building blocks have been defined  in Sec.~\ref{sec:computations} and explicitly computed in appendix~\ref{sec:finite_size_calculations}. Employing the thermodynamic description introduced in section~\ref{sec:bethe_ansatz_solution}, and the pair structure of the excitations \eqref{eq:particle_hole_exc}, it is straightforward to rewrite \eqref{eq:general_time_evolution} in the thermodynamic limit as \cite{dwbc-14,dc-14}
\bea
\lim_{\rm th}\langle \chi^0_N |  \mathcal{O}(t) | \chi^0 _N\rangle &=&  \frac{1}{2}\sum_{n=0}^\infty \frac{1}{n!^2}  \left[\prod_{j=1}^n \int_0^\infty d\mu_j^+ \int_0^\infty d\mu_j^- \rho_{0}(\mu_j^-) \rho_{0}^h(\mu_j^+) e^{- \delta s(\mu_j^+) + \delta s(\mu_j^-) - 2 i t [\delta \omega(\mu_j^+) -\delta \omega(\mu_j^-) ]} \right] \nonumber \\
&\times&\langle {\rho}_{sp} | \mathcal{O} | {\rho}_{sp}, \{ \mu_j^- , - \mu_j^-\to\mu_j^+ , -\mu_j^+ \}_{j=1}^n \rangle  + {\rm mirr.} \,,\
\label{eq:starting_point}
\eea
where 
\bea
{\rm mirr}&=& \frac{1}{2} \sum_{n=0}^\infty \frac{1}{n!^2}  \left[\prod_{j=1}^n \int_0^\infty d\mu_j^+ \int_0^\infty d\mu_j^- \rho_{0}(\mu_j^-) \rho_{0}^h(\mu_j^+) e^{- \delta s(\mu_j^+) + \delta s(\mu_j^-) + 2 i t [\delta \omega(\mu_j^+) -\delta \omega(\mu_j^-)  ]} \right] \nonumber\\
&\times&\langle  {\rho}_{sp}, \{ \mu_j^- , - \mu_j^-\to\mu_j^+ , -\mu_j^+ \}_{j=1}^n  | \mathcal{O} | {\rho}_{sp} \rangle
\label{eq:mirrored_sum}
\eea
corresponds to the second term in the sum \eqref{eq:general_time_evolution} (giving rise to what is sometimes called the mirrored sum). The functions $\rho_0(\mu)$, $\delta \omega(\mu)$ and $\delta s(\mu)$ are defined in \eqref{eq:rho0_function}, \eqref{eq:delta_omega} and \eqref{eq:delta_s} respectively. We also introduced the hole rapidity distribution function of the representative eigenstate, which is related to $\rho_0(\mu)$ by the Bethe equation \eqref{eq:thermo_bethe}. In the limit $c\to\infty$ the latter becomes a simple linear relation, so that one readily obtains
\be
\rho_0^h(\lambda)=\frac{1}{2\pi}-\rho_0(\lambda)=\frac{1}{2\pi}\frac{(\lambda/2)^2}{1+(\lambda/2)^2}\,.
\ee

We can now specify our calculation to the operator $\mathcal{O}=\phi^{\dagger}_{\kappa}(x)\phi_{\kappa}(y)$. Using the explicit expression \eqref{eq:final_ff} for the form factor of the one-body density matrix and the notation \eqref{eq:notation_time_ev}, we can rewrite \eqref{eq:starting_point} as
\bea
\langle  \phi_{\kappa}^{\dagger}(x) \phi_{\kappa}(y) \rangle_t &=& {\rm Det}( 1+ \mathcal{B}_{\kappa}\rho_{0})\nonumber\\
&\times & \frac{1}{2}\sum_{n=0}^{\infty} \frac{1}{n!^2} \left[\prod_{j=1}^n \int_0^\infty d\mu_j^+ \int_0^\infty d\mu_j^-  \rho_{0}(\mu^-_j) \rho_{0}^h(\mu^+_j)       e^{-2i t [\delta \omega(\mu_j^+) -\delta  \omega (\mu_j^-)] - \delta s(\mu_j^+) + \delta s(\mu_j^-) } \right]
\nonumber\\
& \times &
\det_{i,j=1}^{n} \begin{pmatrix}
\mathcal{W}_{\kappa}(\mu^-_i,\mu^+_j) & \mathcal{W}_{\kappa}(\mu^-_i,-\mu^+_j) \\ \mathcal{W}_{\kappa}(-\mu^-_i, \mu^+_j) & \mathcal{W}_{\kappa}(-\mu^-_i, -\mu^+_j) 
\end{pmatrix}
-(\mathcal{B}_{\kappa},\mathcal{W}_{\kappa} \to \mathcal{A}_{\kappa},\mathcal{V}_{\kappa}) + {\rm mirr.}\,,
\eea
where we made use of the compact notation 
\bea
(\mathcal{B}_{\kappa},\mathcal{W}_{\kappa} \to \mathcal{A}_{\kappa},\mathcal{V}_{\kappa})&=& {\rm Det}( 1+ \mathcal{A}_{\kappa}\rho_{0})\nonumber\\
&\times & \frac{1}{2}\sum_{n=0}^{\infty} \frac{1}{n!^2} \left[\prod_{j=1}^n \int_0^\infty d\mu_j^+ \int_0^\infty d\mu_j^-  \rho_{0}(\mu^-_j) \rho_{0}^h(\mu^+_j)       e^{-2i t [\delta \omega(\mu_j^+) -\delta  \omega (\mu_j^-)] - \delta s(\mu_j^+) + \delta s(\mu_j^-) } \right]
\nonumber\\
& \times &
\det_{i,j=1}^{n} \begin{pmatrix}
\mathcal{V}_{\kappa}(\mu^-_i,\mu^+_j) & \mathcal{V}_{\kappa}(\mu^-_i,-\mu^+_j) \\ \mathcal{V}_{\kappa}(-\mu^-_i, \mu^+_j) & \mathcal{V}_{\kappa}(-\mu^-_i, -\mu^+_j) 
\end{pmatrix}\,.
\label{eq:partially_simplified0}
\eea

Next, it follows directly from \eqref{eq:ff_exchange_identity} that
\be
\langle {\rho}_{sp}, \{ \mu_j^- , - \mu_j^-\to\mu_j^+ , -\mu_j^+ \}_{j=1}^n\}_{j=1}^n|  \phi_{\kappa}^{\dagger}(x) \phi_{\kappa}(y)  | {\rho}_{sp}  \rangle
 =\langle {\rho}_{sp} |  \phi_{\kappa}^{\dagger}(x) \phi_{\kappa}(y)  | {\rho}_{sp}, \{ \mu_j^- , - \mu_j^-\to\mu_j^+ , -\mu_j^+ \}_{j=1}^n \rangle\,,
\label{eq:thermodyn_property_ff}
\ee
which allows us to show that for $\mathcal{O}=\phi_{\kappa}^{\dagger}(x)\phi_{\kappa}(y)$ the mirrored sum \eqref{eq:mirrored_sum} coincides with the first term in the r.h.s. of \eqref{eq:starting_point}. Indeed, using \eqref{eq:thermodyn_property_ff}, we have
\bea
{\rm mirr}&=& \frac{1}{2} \sum_{n=0}^\infty \frac{1}{n!^2}  \left[\prod_{j=1}^n \int_0^\infty d\mu_j^+ \int_0^\infty d\mu_j^- \rho_{0}(\mu_j^-) \rho_{0}^h(\mu_j^+) e^{- \delta s(\mu_j^+) + \delta s(\mu_j^-) + 2 i t [\delta \omega(\mu_j^+) -\delta \omega(\mu_j^-) ]} \right] \nonumber\\
&\times &\langle {\rho}_{sp} |  \phi_{\kappa}^{\dagger}(x) \phi_{\kappa}(y)  | {\rho}_{sp}, \{ \mu_j^- , - \mu_j^-\to\mu_j^+ , -\mu_j^+ \}_{j=1}^n \rangle  
\nonumber\\
&= &{\rm Det}( 1+ \mathcal{B}_{\kappa}\rho_{0})\nonumber\\
&\times & \frac{1}{2}\sum_{n=0}^{\infty} \frac{1}{n!^2} \left[\prod_{j=1}^n \int_0^\infty d\mu_j^+ \int_0^\infty d\mu_j^-  \rho_{0}(\mu^-_j) \rho_{0}^h(\mu^+_j)       e^{+2i t [\delta \omega(\mu_j^+) -\delta  \omega (\mu_j^-)] - \delta s(\mu_j^+) + \delta s(\mu_j^-) } \right]
\nonumber\\
& \times &
\det_{i,j=1}^{n} \begin{pmatrix}
\mathcal{W}_{\kappa}(\mu^-_i,\mu^+_j) & \mathcal{W}_{\kappa}(\mu^-_i,-\mu^+_j) \\ \mathcal{W}_{\kappa}(-\mu^-_i, \mu^+_j) & \mathcal{W}_{\kappa}(-\mu^-_i, -\mu^+_j)
\end{pmatrix}
-(\mathcal{B}_{\kappa},\mathcal{W}_{\kappa} \to \mathcal{A}_{\kappa},\mathcal{V}_{\kappa})\nonumber \\
&= &{\rm Det}( 1+ \mathcal{B}_{\kappa}\rho_{sp})\nonumber\\
&\times & \frac{1}{2}\sum_{n=0}^{\infty} \frac{1}{n!^2} \left[\prod_{j=1}^n \int_0^\infty d\mu_j^+ \int_0^\infty d\mu_j^-  \rho_{0}(\mu^+_j) \rho_{0}^h(\mu^-_j)  e^{+2i t [\delta \omega(\mu_j^-) -\delta  \omega (\mu_j^+)] - \delta s(\mu_j^-) + \delta s(\mu_j^+) } \right]
\nonumber\\
& \times &
\det_{i,j=1}^{n} \begin{pmatrix}
\mathcal{W}_{\kappa}(\mu^+_i,\mu^-_j) & \mathcal{W}_{\kappa}(\mu^+_i,-\mu^-_j) \\ \mathcal{W}_{\kappa}(-\mu^+_i, \mu^-_j) & \mathcal{W}_{\kappa}(-\mu^+_i, -\mu^-_j) 
\end{pmatrix}
-(\mathcal{B}_{\kappa},\mathcal{W}_{\kappa} \to \mathcal{A}_{\kappa},\mathcal{V}_{\kappa})\,,
\label{eq:long_equation}
\eea
where in the last step we have relabeled the integration variables. We can now make use of the identities
\bea
\rho_{0}^h(\lambda)e^{-\delta s(\lambda)}&=&\rho_{0}(\lambda)e^{\delta s(\lambda)}\,,
\label{eq:identity1}
\eea
and
\bea
\det_{i,j=1}^{n} \begin{pmatrix}
\mathcal{W}_{\kappa}(\mu^+_i,\mu^-_j) & \mathcal{W}_{\kappa}(\mu^+_i,-\mu^-_j) \\ \mathcal{W}_{\kappa}(-\mu^+_i, \mu^-_j) & \mathcal{W}_{\kappa}(-\mu^+_i, -\mu^-_j) 
\end{pmatrix}&=&\det_{i,j=1}^{n}\left[ \left(\begin{array}{cc}
\mathcal{W}_{\kappa}(\mu^+_i,\mu^-_j) & \mathcal{W}_{\kappa}(\mu^+_i,-\mu^-_j) \\ \mathcal{W}_{\kappa}(-\mu^+_i, \mu^-_j) & \mathcal{W}_{\kappa}(-\mu^+_i, -\mu^-_j) 
\end{array}\right)^{T}\ \right]\nonumber\\
&=&\det_{i,j=1}^{n} \begin{pmatrix}
\mathcal{W}_{\kappa}(\mu^-_i,\mu^+_j) & \mathcal{W}_{\kappa}(\mu^-_i,-\mu^+_j) \\ \mathcal{W}_{\kappa}(-\mu^-_i, \mu^+_j) & \mathcal{W}_{\kappa}(-\mu^-_i, -\mu^+_j) 
\end{pmatrix}\,,
\label{eq:identity2}
\eea
where $A^T$ denotes the transpose of the matrix $A$, while the last equality follows from $\mathcal{W}_{\kappa}(u,v)=\mathcal{W}_{\kappa}(v,u)$ (this is true since the kernel $\mathcal{B}_{\kappa}(u,v)$  is symmetric under exchange $u\leftrightarrow v$). An analogous calculation can be carried out for the kernel $\mathcal{V}_{\kappa}$. Plugging Eqs.~\eqref{eq:identity1} and \eqref{eq:identity2} into \eqref{eq:long_equation}, one immediately has that for $\mathcal{O}=\phi_{\kappa}^{\dagger}(x)\phi_{\kappa}(y)$ the mirrored sum \eqref{eq:mirrored_sum} coincides with the first term in the r.h.s. of \eqref{eq:starting_point}. Summarizing, one can rewrite \eqref{eq:partially_simplified0} in terms of a single sum, as
\bea
\langle  \phi_{\kappa}^{\dagger}(x) \phi_{\kappa}(y) \rangle_t &=& {\rm Det}( 1+ \mathcal{B}_{\kappa}\rho_{0})\nonumber\\
&\times & \sum_{n=0}^{\infty} \frac{1}{n!^2} \left[\prod_{j=1}^n \int_0^\infty d\mu_j^+ \int_0^\infty d\mu_j^-  \rho_{0}(\mu^-_j) \rho_{0}^h(\mu^+_j)       e^{-2i t [\delta \omega(\mu_j^+) -\delta  \omega (\mu_j^-)] - \delta s(\mu_j^+) + \delta s(\mu_j^-) } \right]
\nonumber\\
& \times &
\det_{i,j=1}^{n} \begin{pmatrix}
\mathcal{W}_{\kappa}(\mu^-_i,\mu^+_j) & \mathcal{W}_{\kappa}(\mu^-_i,-\mu^+_j) \\ \mathcal{W}_{\kappa}(-\mu^-_i, \mu^+_j) & \mathcal{W}_{\kappa}(-\mu^-_i, -\mu^+_j) 
\end{pmatrix}
-(\mathcal{B}_{\kappa},\mathcal{W}_{\kappa} \to \mathcal{A}_{\kappa},\mathcal{V}_{\kappa}) \,.
\label{eq:partially_simplified1}
\eea

Following \cite{dc-14}, we now note that for a measure $\mu(y)$ which is well-defined on a domain $X \subset \mathbb{R}$, one can use the identity \cite{Mehta-04}
\be
\Big[ \prod_{\alpha=1}^n \int_X d\mu(y_\alpha) \Big] \det_{\alpha=[1,2n],\beta=[1,n]} \Big( A_\alpha(y_\beta)\ B_\alpha(y_\beta) \Big) = n! \sqrt{\det_{\alpha,\beta=1}^{2n} a_{\alpha \beta}}= n! \text{Pf}\left[ a_{\alpha, \beta}\right]_{\alpha,\beta=1}^{2n} \,.
\label{eq:pfaffian_identity}
\ee
Here we introduced the Pfaffian $\text{Pf}\left[ a_{\alpha, \beta}\right]_{\alpha,\beta=1}^{2n} $ of the matrix $a_{\alpha,\beta}$, which is defined as
\be
a_{\alpha, \beta}= \int_X d\mu(y) \Big( A_\alpha(y)B_\beta(y) -    A_\beta(y)B_\alpha(y) \Big) \,.
\ee
In order to apply \eqref{eq:pfaffian_identity} to \eqref{eq:partially_simplified1}, we can choose $A_\alpha(y_\beta) = \mathcal{W}_{\kappa}(\mu_{\alpha}^-, y_{\beta} )$ for $\alpha=1,\ldots n$ and $A_\alpha(y_\beta) = \mathcal{W}_{\kappa}(-\mu_{\alpha-n}^-, y_{\beta} )$ for $\alpha= n+1,\ldots 2n$. Analogously, we can set $B_\alpha(y_\beta) = \mathcal{W}_{\kappa}(\mu^-_{\alpha}, - y_{\beta})$ for $\alpha=1,\ldots n$ and $B_\alpha(y_\beta) = \mathcal{W}_{\kappa}(-\mu_{\alpha-n}^-,- y_{\beta} )$  for $\alpha= n+1,\ldots 2n$. Then we can apply \eqref{eq:pfaffian_identity} to \eqref{eq:partially_simplified1} for each $n$: in particular, making use of the definition \eqref{eq:phi_plus} we can perform the integration over the rapidities $\mu_j^+$, obtaining
\bea
&\ &\left[\prod_{j=1}^n \int_0^\infty d\mu_j^{+}  \varphi_{+}^{(t)}(\mu_j^+)\right] \det_{i,j=1}^{n}
\begin{pmatrix}
\mathcal{W}_{\kappa}(\mu^-_i,\mu^+_j) & \mathcal{W}_{\kappa}(\mu^-_i,-\mu^+_j) \\ \mathcal{W}_{\kappa}(-\mu^-_i, \mu^+_j) & \mathcal{W}_{\kappa}(-\mu^-_i, - \mu^+_j) 
\end{pmatrix} \nonumber \\
&\ &\hspace{0.5cm}= n! {\rm Pf}\left[ \begin{pmatrix}
\Theta^{\mathcal{W}}(\mu^-_i,\mu^-_j) & \Theta^{\mathcal{W}}(\mu^-_i,-\mu^-_j) \\ \Theta^{\mathcal{W}}(-\mu^-_i, \mu^-_j) & \Theta^{\mathcal{W}}(-\mu^-_i, - \mu^-_j) 
\end{pmatrix}\right]_{i,j=1}^{n} \,,
\eea
where we introduced
\bea
\Theta^{\mathcal{W}}(u,v)&=&\int_{-\infty}^{+\infty}dy \varphi^{(t)}_+(y)\mathcal{W}_{\kappa}(u,y)\mathcal{W}_{\kappa}(v,-y)\,.\label{eq:W_kernel}
\eea
In the last step, we explicitly used the antisymmetry of the function $\varphi^{(t)}_{+}(\mu)$, namely $\varphi^{(t)}_{+}(-\mu)=-\varphi^{(t)}_{+}(\mu)$.  An analogous computation can be carried out for the kernel $\mathcal{V}_{\kappa}$, leading to the corresponding definition 
\bea
\Theta^{\mathcal{V}}(u,v)&=&\int_{-\infty}^{+\infty}dy \varphi^{(t)}_+(y)\mathcal{V}_{\kappa}(u,y)\mathcal{V}_{\kappa}(v,-y)\,.\label{eq:V_kernel}
\eea

Collecting all the previous calculations, we can recast \eqref{eq:partially_simplified1} into the simplified form
\bea
\langle  \phi_{\kappa}^{\dagger}(x) \phi_{\kappa}(y) \rangle_t &=& \text{Det}( 1+ \mathcal{B}_{\kappa} \rho_{0}) \sum_{n=0}^{\infty} \frac{1}{n!} \left[ \prod_{j=1}^n \int_{0}^{+ \infty}  d\mu_j^-    \varphi_{-}^{(t)}(\mu_j^-) \right]\nonumber\\
&\times &  \text{Pf}
\left[ \begin{pmatrix}
\Theta^{\mathcal{W}}(\mu^-_i,\mu^-_j) & \Theta^{\mathcal{W}}(\mu^-_i,-\mu^-_j) \\ \Theta^{\mathcal{W}}(-\mu^-_i,  \mu^-_j) & \Theta^{\mathcal{W}}(-\mu^-_i, - \mu^-_j) \end{pmatrix}\right]_{i,j=1}^{n} - (\mathcal{B}_{\kappa},\Theta^{\mathcal{W}} \to \mathcal{A}_{\kappa},\Theta^{\mathcal{V}})\,,
\label{eq:partially_simplified2}
\eea
where $\varphi^{(t)}_{-}(\mu)$ is defined in \eqref{eq:phi_minus}.
The series appearing in \eqref{eq:partially_simplified2} can be summed explicitly, yielding a final expression written in terms of Fredholm Pfaffians, namely
\bea
\sum_{n=0}^{\infty} \frac{1}{n!} \left[\prod_{j=1}^n \int_{0}^{+ \infty}  d\mu_j^-  \varphi_{-}^{(t)}(\mu_j^-) \right] 
{\rm Pf}\left[ \begin{pmatrix}
\Theta^{\mathcal{W}}(\mu^-_i,\mu^-_j) & \Theta^{\mathcal{W}}(\mu^-_i,-\mu^-_j) \\ \Theta^{\mathcal{W}}(-\mu^-_i,  \mu^-_j) & \Theta^{\mathcal{W}}(-\mu^-_i, - \mu^-_j) 
\end{pmatrix}\right]_{i,j=1}^{n} = \text{Pf}\Big( \boldsymbol{J} +P_0 \boldsymbol{\Theta}^{\mathcal{W}} \varphi_{-}^{(t)}  P_0\Big) \,,
\eea
where we introduced the $2 \times 2$ matrices
\bea
\boldsymbol{\Theta}^{\mathcal{W}} &=& \begin{pmatrix}
 \Theta^{\mathcal{W}}_{++} &   \Theta^{\mathcal{W}}_{+-}  \\  \Theta^{\mathcal{W}}_{-+}&  \Theta^{\mathcal{W}}_{--}
\end{pmatrix}\,,\\ 
 \boldsymbol{J} &=& \begin{pmatrix}
 0  & 1 \\ -1 & 0
\end{pmatrix} \,.
\eea
As usual, we indicated with $1$ the identity operator, while we defined the kernels $\Theta^{\mathcal{W}}_{\pm \pm}(\lambda,\mu)=\Theta^{\mathcal{W}}(\pm\lambda,\pm\mu)$, together with the projector $P_0$ on the positive real line $x>0$. The Fredholm Pfaffian is a mathematical object which can be related to the Fredholm determinant through the relation
\be
{\rm Pf}(\mathbf{J}+\mathbf{\Theta})^2={\rm Det}\left(\mathbf{I}-\mathbf{J}\mathbf{\Theta}\right)\,.
\label{eq:pfaffain_fredholm}
\ee
Details about Fredholm Pfaffians can be found in  \cite{DoCa12,BoKa07}. Here we just need Eq.~\eqref{eq:pfaffain_fredholm} as a
woking formula.

Using \eqref{eq:pfaffain_fredholm} we can rewrite
\bea
{\rm Pf}\left(\mathbf{J}+P_0\mathbf{\Theta}^{\mathcal{W}}\varphi^{(t)}_- P_0\right)=\sqrt{{\rm Det}\left(\mathbf{I}-P_0\mathbf{J}\mathbf{\Theta}^{\mathcal{W}} \varphi_-^{(t)}P_0\right)}\,.
\label{eq:sqrt_determinant}
\eea
In this equation there is an ambiguity on the sign of the square root of the Fredholm determinant which 
can be fixed by requiring a few properties on the time-dependent one-body density matrix. 
This is briefly mentioned in the main text after Eq.~\eqref{eq:final_result}, and discussed in detail in appendix~\ref{sec:fredholm}.

From the antisymmetry of the function $ \varphi_{-}^{(t)}(y)$ defined in \eqref{eq:phi_minus}, and Sylvester's determinant identity
\be
{\rm Det}\left(1+AB\right)={\rm Det}\left(1+BA\right)\,,
\ee
one can show
\be
{\rm Det}(\boldsymbol{I} -P_0 \boldsymbol{J}\mathbf{\Theta}^{\mathcal{W}} \varphi_{-}^{(t)}P_0 ) = {\rm Det}
 \begin{pmatrix}
 1 - P_0  \varphi_{-}^{(t)} \Theta^{\mathcal{W}}_{-+} P_0  &  - P_0  \varphi_{-}^{(t)}  \Theta^{\mathcal{W}}_{--} P_0  \\  P_0  \varphi_{-}^{(t)}   \Theta^{\mathcal{W}}_{++} P_0  & 1 + P_0 \varphi_{-}^{(t)}   \Theta^{\mathcal{W}}_{+-} P_0 
\end{pmatrix}  = {\rm Det}( 1+ \varphi_{-}^{(t)}   \Theta^{\mathcal{W}}_{+-} ) \,
\label{eq:fredholm_whole_real_axis}\,,
\ee
where the last Fredholm determinant is defined on the whole real axis. The validity of \eqref{eq:fredholm_whole_real_axis} is immediately proven taking finite dimensional representations of the Fredholm determinant, and exploiting the properties of matrix determinants under exchange of rows and columns.

At this point, one technical complication arises w. r. t. the bosonic case treated in \cite{dc-14}. In fact, for $\kappa=0$ one has
\bea
\mathcal{A}_{\kappa=0}(-\lambda,-\mu)=\mathcal{A}_{\kappa=0}(\lambda,\mu)\,,\\
\mathcal{B}_{\kappa=0}(-\lambda,-\mu)=\mathcal{B}^{\ast}_{\kappa=0}(\lambda,\mu)\,.
\eea
These relations were explicitly used in the derivation of \cite{dc-14} and are no longer valid for $\kappa\neq 0$. Luckily, this issue can be easily overcome as follows. We consider the case of the kernel $\mathcal{A}_{\kappa}$. Since we won't use any specific property of the latter, the same derivation holds also for the kernel $\mathcal{B}_{\kappa}$. Define the singular kernel
\be
\Delta(u,v)=\delta(u+v)\,,
\ee
so that
\be
\Theta^{\mathcal{V}}_{+-}(u,v)=\Theta^{\mathcal{V}}(u,-v)=\left[\mathcal{V}_{\kappa}\cdot\varphi_+^{(t)}\cdot\Delta \cdot\mathcal{V}_{\kappa}\cdot\Delta \right](u,v)=\left[(1+\mathcal{A}_{\kappa}\rho_0)^{-1}\mathcal{A}_{\kappa} \varphi^{(t)}_+\Delta(1+\mathcal{A}_{\kappa}\rho_0)^{-1}\mathcal{A}_{\kappa}\Delta\right](u,v)\,,
\ee
where we used
\be
\mathcal{V}_{\kappa}=(1+\mathcal{A}_{\kappa}\rho_0)^{-1}\mathcal{A}_{\kappa}\,.
\ee
Then, using once again Sylvester's determinant identity we have
\bea
{\rm Det}\left(1+\varphi^{(t)}_-\Theta^{\mathcal{V}}_{+-}\right)&=&{\rm Det}\left(1+\Theta^{\mathcal{V}}_{+-}\varphi^{(t)}_-\right)=\frac{{\rm Det}\left(1+\mathcal{A}_{\kappa}\rho_0+\mathcal{A}_{\kappa}\varphi_+^{(t)}\Delta(1+\mathcal{A}_{\kappa}\rho_0)^{-1}\mathcal{A}_{\kappa}\Delta\varphi_-^{(t)}\right)}{{\rm Det}\left(1+\mathcal{A}_{\kappa}\rho_0\right)}\nonumber\\
&=&\frac{{\rm Det}\begin{pmatrix}
1+\mathcal{A}_{\kappa}\rho_0 & -\mathcal{A}_{\kappa}\varphi^{(t)}_+\Delta\\
\mathcal{A}_{\kappa}\Delta \varphi^{(t)}_- & 1+\mathcal{A}_{\kappa}\rho_0
\end{pmatrix}}{{\rm Det}\left(1+\mathcal{A}_{\kappa}\rho_0\right)^2}\,,
\label{eq:intermediate_step_determinants}
\eea
where we used the identity
\be
{\rm Det}\begin{pmatrix}
A & B\\
C & D
\end{pmatrix}= {\rm Det} D\times {\rm Det} (A-BD^{-1}C)\,.
\ee
We now note 
\bea
[\mathcal{A}_{\kappa}\Delta](u,v)&=&\mathcal{A}_{\kappa}(u,-v)=\mathcal{A}_{\kappa}^{+-}(u,v)\,,\\
\ [\mathcal{A}_{\kappa}\varphi_+^{(t)}\Delta](u,v)&=&\mathcal{A}_{\kappa}(u,-v)\varphi_+^{(t)}(-v)=-\mathcal{A}_{\kappa}(u,-v)\varphi_+^{(t)}(v)=-\mathcal{A}_{\kappa}^{+-}(u,v)\varphi_+^{(t)}(v)\,,
\eea
where in the last step we used $\varphi^{(t)}_+(-v)=-\varphi^{(t)}_+(v)$, and where $\mathcal{A}_{\kappa}^{+-}(u,v)$ is defined in \eqref{eq:A_plusminus}.

Putting everything together, one immediately obtains the final result \eqref{eq:final_result} presented in the main text.

\twocolumngrid

\end{document}